\documentclass[aps, prl, amsmath, superscriptaddress,preprintnumbers,twocolumn]{revtex4-2}
\usepackage{graphicx}
\usepackage{dcolumn}
\usepackage{bm}
\usepackage{amsmath}
\usepackage{subfigure}
\usepackage{amsfonts}
\usepackage[svgnames]{xcolor}
\usepackage{longtable}
\usepackage{appendix}
\usepackage{multirow}
\usepackage{booktabs}
\usepackage{tabularx}
\usepackage[colorlinks,
            linkcolor=blue,
			filecolor=blue,
			urlcolor= blue,
			citecolor=blue,
            ]{hyperref}
\usepackage{amsthm}
\usepackage{setspace}
\usepackage[T1]{fontenc}
\usepackage{extarrows}
\usepackage{lineno}

\usepackage[percent]{overpic}
\usepackage{dsfont}
\usepackage{floatrow}
\usepackage{physics}

\newcommand{\kett}[1]{\left.\ket{#1}\right\rangle}
\newcommand{\braa}[1]{\left\langle\bra{#1}\right.}
\newcommand{\brakett}[2]{\left\langle\braket{#1}{#2}\right\rangle}
\newcommand{\ee}{\mathrm{e}}

\newcommand{\q}{{\mathrm{p}}}
\newcommand{\ii}{\mathrm{i}}

\newcommand{\phii}[0]{\hat{\Phi}}
\newcommand{\mm}[0]{\hat{\mathcal{M}}}

\newcommand{\bij}{{\beta_{ij}}}

\newcommand{\blue}[1]{\textcolor{blue}{#1}}

\begin{document}
%\begin{CJK*}{UTF8}{gbsn}
%\linenumbers

%\title{Non-Hermitian Physics in Quantum Channels: Pseudo-Hermiticity, Spectrum Estimation and Exceptional Points}
%\title{Non-Hermitian Physics in Quantum Channels: Pseudo-Hermiticity, Spectrum Measurement and Application to Hamiltonian Parameter Estimation}
\title{{Spectrum measurement of quantum channels and application to Hamiltonian parameter estimation}}
\author{Yuan-De Jin}
\affiliation{State Key Laboratory of Semiconductor Physics and Chip Technologies, Institute of Semiconductors, Chinese Academy of Sciences, Beijing 100083, China}
\affiliation{Center of Materials Science and Opto-Electronic Technology, University of Chinese Academy of Sciences, Beijing 100049, China}

\author{Wen-Long Ma}
\email{wenlongma@semi.ac.cn}
\affiliation{State Key Laboratory of Semiconductor Physics and Chip Technologies, Institute of Semiconductors, Chinese Academy of Sciences, Beijing 100083, China}
\affiliation{Center of Materials Science and Opto-Electronic Technology, University of Chinese Academy of Sciences, Beijing 100049, China}
\date{\today }
\begin{abstract}
Quantum channels describe the most general {dynamics} of open quantum systems. {A quantum channel, as a linear map on vectorized quantum states, can be represented by a single matrix, whose spectrum is called the channel spectrum. Here we propose a general method to measure the channel spectrum and apply this method to Hamiltonian parameter estimation.} We first demonstrate that the channel spectrum can be measured by tracking the probability of a specific outcome in repeated application of the same channel. Then we construct and analyze {a class of concatenated channels, with each one being a unitary channel followed by a weak-measurement channel induced by a Ramsey sequence of a probe qubit}. We show that the spectrum measurement of such concatenated channels can be utilized for estimating the parameters in the free Hamiltonians generating the unitary channels of the target system. As practical examples, we numerically demonstrate that a probe spin qubit can accurately sense nuclear spin clusters for nanoscale nuclear magnetic resonance.
%We first demonstrate that any quantum channel is a pseudo-Hermitian matrix in this representation if it is diagonalizable with a discrete spectrum because its eigenvalues are either real or in complex conjugate pairs. Regarding this property, 
%generated by a free Hamiltonian on a target quantum system

\end{abstract}

\maketitle

\textit{Introduction.} %{\color{red} Focus on the generalization of non-Hermitian physics from Hamiltonians to open quantum dynamics. Review the existing works on Lindbladian dynamics and stress that our paper goes a step further to general quantum channels. Moreover, our results are relevant to existing experiments.}
%Non-Hermitian (NH) physics originates from the study of nonconservative dynamics in classical and quantum open systems. It can describe various nonconservative phenomena, such as photon loss, friction, dissipation, and quantum measurement backaction, through effective NH Hamiltonians. Various effects and applications have been found in NH systems with parity–time (PT) symmetry or more general pseudo-Hermiticity, such as NH skin effect, NH bulk–boundary correspondence, and sensing enhanced by exceptional points (EPs). 
%Non-Hermitian (NH) physics originates from the study of nonconservative dynamics in open classical and quantum systems \cite{ashidaNonHermitianPhysics2020}. 
{Open classical and quantum systems often exhibit nonconservative phenomena  \cite{rivas2012,ashidaNonHermitianPhysics2020}}, such as photon gain and loss in photonics \cite{wang2021,wangNonHermitianOpticsPhotonics2023,takata2018,ozdemir2019}, friction in mechanics \cite{yoshida2019,huber2016}, dissipation in open quantum systems \cite{rotter2009,li2019}, and backaction in quantum measurement \cite{clerk2010,wiseman2009}. {While effective non-Hermitian (NH) Hamiltonians can provide a conceptually simple and intuitive approach to understand these nonconservative processes  \cite{Malzard2015,Yokomizo2019,Rivero2020,Yao2018,Gong2018,bergholtz2021,ding2022,okuma2023,Halder2024,Hamazaki2019,Kawabata2019,Hamazaki2020,Kawabata2022,Garc2022}, they cannot fully describe the evolution of open quantum systems.} For example, the Lindblad quantum dynamics can be unraveled into stochastic quantum trajectories \cite{longhi2020,gneiting2022,donvil2023}, often with the particular no-jump branch described by effective NH Hamiltonians. Then it is necessary to fully study the Lindbladian (or more general Liouvillian) generators of open quantum dynamics \cite{roccati2022,gorini1976,lindblad1976}. Such generators are often NH matrices whose spectra govern relaxation modes and steady states \cite{Albert2014,Nakagawa2021,Popkov2021,Marche2025,Chen2025}.

As the Lindblad formalism describes only a special class of open quantum dynamics, {the most general dynamics of an open quantum system is described by completely positive and trace-preserving maps}, also called quantum channels \cite{wolf2012,watrous2018,Nielsen2010,Caruso2014,Bidhi2025}. Besides continuous-time Lindbladian dynamics, quantum channels can also {conveniently} describe indivisible or intrinsically discrete-time open quantum dynamics, such as non-Markovian quantum dynamics \cite{Breuer2016,deVega2017}, quantum collision dynamics \cite{ciccarello2022,Scarani2002}, noisy quantum gates \cite{Gu2023}, quantum error correction or recovery operations \cite{Gilyen2022,Zheng2025}, and sequential quantum measurement and control processes \cite{blok2014manipulating,cujia2019,pfender2019,Nakagawa2025}. A quantum channel is often a NH operator on vectorized density operators \cite{watrous2018}, whose {spectral properties {relevant} to physical applications} remain largely unexplored.

Quantum channels also provide a general approach to study quantum trajectories and measurement statistics. Specifically, it can always be decomposed as a set of completely-positive operations, with each operation corresponding to a measurement outcome after a dilated unitary operation on an enlarged system containing the system and an environment \cite{Caruso2014}. Recently, the measurement statistics in repetitive quantum channels has been utilized for tracking the precession of single nuclear spins in nanoscale nuclear magnetic resonance (NMR) \cite{cujia2019,pfender2019}, however, {there is still a lack of a general framework to describe these phenomena}. To develop such a framework can help answer the following questions: What is the intrinsic relation between {spectral properties} of a channel and the measurement statistics in repetitive channels? How can the measurement statistics information be useful for quantum sensing or quantum system learning? Can we construct {a general class of channels} to demonstrate such applications?

% In this paper, we reveal the rich NH physics in quantum channels and explore its applications. We first prove that the matrix of a quantum channel is pseudo-Hermitian if it is diagonalizable, due to its special spectral structure. The main result is to build the connection between the spectral structure of a quantum channel and the measurement statistics in {repeated applications of the channel, specifically, the channel spectrum can be efficiently measured by tracking the probability of a specific outcome in the repeated channel}. We further construct a typical class of quantum channels, each composed of a unitary channel and a weak-measurement channel, whose spectral properties can be well captured by perturbation theory. Through practical examples, we verify that the spectrum measurement of such concatenated channels enables us to learn the free Hamiltonians generating the unitary channel. These results provide a general framework for recent experiments in tracking single nuclear spins with sequential weak measurements \cite{cujia2019,pfender2019}, such that these schemes can be generalized to detect arbitrary nuclear spin clusters in nanoscale NMR.  

In this Letter, we present a general framework to measure the quantum channel spectrum and explore its application to Hamiltonian parameter estimation. The main result is to build the connection between the {channel spectrum} and the measurement statistics in repeated applications of the same channel. Specifically, the channel spectrum can be efficiently measured by tracking the probability of a specific outcome in repetitive quantum channels. We further construct a typical class of quantum channels, each composed of a unitary channel and a weak-measurement channel, whose spectral properties can be well captured by perturbation theory. Through practical examples, we verify that the spectrum measurement of such a concatenated channel enables us to learn the parameters in the free Hamiltonian generating the unitary channel. These results provide a general model to understand recent experiments in tracking the precession of single nuclear spins with repetitive weak measurements \cite{cujia2019,pfender2019}, such that these schemes can be generalized to detect arbitrary nuclear spin clusters for nanoscale NMR.

\textit{Preliminaries and spectral properties of quantum channels.}  We start by introducing the basic notations in defining different representations of a quantum channel \cite{watrous2018}.
%The Stinespring representation of a quantum channel describes the evolution of a system state $\rho$ by first coupling the system to a quantum environment and then partially tracing over the environment, i.e., $\Phi(\rho)=\Tr_{E}[U_{\rm tot}(\rho_{E}\otimes\rho) U^\dagger_{\rm tot}]$, where the environment can always be enlarged so that $\rho_{E}=|\phi\rangle_E\langle \phi|$ is a pure state, $U_{\rm tot}$ is a unitary transformation of the composite system, $(\cdot)^{\dagger}$ denotes the Hermitian conjugation, and $\Tr_{\rm E}$ is the partial trace over the environment.After tracing over the environment basis $\{|e_j\rangle_E\}_{j=1}^r$, 
{The Kraus representation of a quantum channel is}
\begin{equation}\label{Eq:Kraus}
  \Phi(\rho)=\sum_{i=1}^r\mathcal{M}_i(\rho)=\sum_{i=1}^r M_i \rho M_i^\dagger,
\end{equation}
where $\{M_i\}_{i=1}^r$ is a set of Kraus operators %with $M_i=\langle e_j|U_{\rm tot}|\phi\rangle_E$, 
satisfying $\sum_{i=1}^r M_i^\dagger M_i=\mathbb{I}$ with $\mathbb{I}$ being the identity operator. Since a channel is a superoperator (i.e., a linear map acting on operators rather than vectors) on the $d$-dimensional Hilbert space, to analyze the spectrum of $\Phi$, it is convenient to represent it by a single matrix acting on the $d^2$-dimensional {vectorized operator space}. In such a space, an operator $R=\sum_{m,n=1}^d R_{mn}|m\rangle\langle n|$ is vectorized as $\kett{R} = \sum_{m,n=1}^d R_{mn} \, |m\rangle \otimes |n\rangle$ and the inner product is defined as $\brakett{L}{R}=\Tr(L^{\dagger}R)$, called the Hilbert-Schmidt scalar product. Then a superoperator $X(\cdot)Y$ is equivalent to an operator $X\otimes Y^T$ on the vectorized operator space, where $X$, $Y$ are arbitrary operators {on the $d$-dimensional Hilbert space} and $(\cdot)^T$ denotes the matrix transposition. Then in the vectorized operator space, $\Phi$ can be represented as $\phii=\sum_{i=1}^r M_i \otimes M_i^*$ with $(\cdot)^*$ denoting the matrix conjugation. Note that we add hats for operators on the vectorized operator space.

\begin{figure}[htbp]
\centering
  \includegraphics[width=8.7cm]{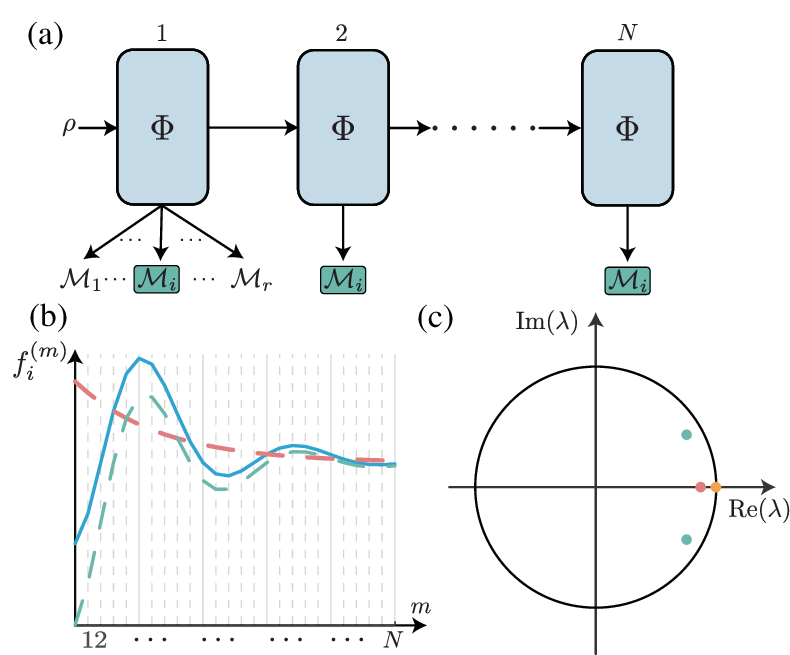}
  \caption{Schematic illustration of quantum channel spectrum measurement. (a) The system evolves with $N$ repetitive quantum channels. Each channel has $r$ measurement outcomes, and we track the frequency $f_i^{(m)}$ of a particular outcome $i$ in the $m$th channel to measure the channel spectrum. (b) Schematic of $f_i^{(m)}$ (blue solid line) as a function of measurement cycle number $m$. A real eigenvalue of the channel contributes an exponential decay (dashed red line), while a pair of complex conjugate eigenvalues contribute a damped oscillation (dashed green line).  (c) The channel spectrum can be accurately inferred from $\{f_i^{(m)}\}_{m=1}^N$ by the MP method. 
  }\label{Fig1}
\end{figure}

We then introduce the spectral properties of quantum channels. A quantum channel can be spectrally decomposed in the vectorized operator space if it is diagonalizable 
\begin{equation}\label{Eq:Phi}
 \phii=\sum_{j=1}^{d^2} \lambda_j\kett{R_j}\braa{L_j},
\end{equation}
where {$\{\kett{R_j},\kett{L_j}\}$ is a complete biorthonormal system satisfying $\brakett{L_i}{R_j}=\delta_{ij}$} with $\delta_{ij}$ being the Kronecker delta, and all the eigenvalues $\{\lambda_j\}$ are within a unit circle of the complex plane \cite{wolf2012}. {Since the quantum channel is a Hermitian-preserving map, i.e., $\Phi(R_k)^{\dagger}=\Phi(R_k^{\dagger})$, we have $\Phi(R_k^{\dagger})=\lambda_k^*R_k^{\dagger}$ with $R_k^{\dagger}$ ($L_k^{\dagger}$) becoming the right (left) eigenstate for eigenvalue $\lambda_k^*$. Thus the eigenvalues are either real or in complex conjugate pairs. This leads to an interesting property of the quantum channel, that is, $\phii$ is pseudo-Hermitian if it is diagonalizable. }

{Recall that an operator $H$ is pseudo-Hermitian if there exists an invertible Hermitian metric operator $\eta$ such that $H^\dagger=\eta H \eta^{-1}$ \cite{mostafazadeh2010}}. A theorem in Ref. \cite{mostafazadeh2002} says that a linear operator acting on the Hilbert space with a complete biorthonormal system and a discrete spectrum is pseudo-Hermitian if and only if its spectrum is either entirely real or in complex conjugate pairs with the same degeneracy. One can see that the channel spectrum exactly satisfies this condition. {We can also explicitly construct a Hermitian metric operator 
$
  \eta=\sum_{\{j|\lambda_j\in\mathbb{R}\}}\kett{L_j}\braa{L_j}+\sum_{\{k|\lambda_k\in\mathbb{C}\textbackslash \mathbb{R}\}}\kett{L_k}\langle\langle L_k^{\dagger}|
$, and its inverse
$
  \eta^{-1}=\sum_{\{j|\lambda_j\in\mathbb{R}\}}\kett{R_j}\braa{R_j}+\sum_{\{k|\lambda_k\in\mathbb{C}\textbackslash \mathbb{R}\}}|R_k^{\dagger}\rangle\rangle \braa{R_k}
$. } Then $\eta\phii\eta^{-1}=\phii^\dagger$,
 where $\phii^\dagger=\sum_{i=1}^r M_i^{\dagger} \otimes M_i^{T}$ corresponds to the dual channel  of $\Phi$, i.e., $\Phi^{\dagger}(\cdot)=\sum_i M_i^{\dagger}(\cdot)M_i$. Thus, $\phii$ is pseudo-Hermitian if it is diagonalizable. We note that previous works have found the pseudo-Hermiticity of Lindbladians \cite{Stenholm2002,Jakob2003,Stenholm2004}, and we can further prove that any Hermitian-preserving map on the operator space is pseudo-Hermitian if it is diagonalizable (see Sec. S1 of the Supplementary Material (SM) \footnote{See the Supplementary Material \url{http://link.aps.org/supplemental/10.1103/66yg-qs8x} for the proof on the pseudo-Hermiticity of Hermitian-preserving maps, measurement statistics for non-diagonalizable quantum channels and correlation measurements, and details in Hamiltonian parameter estimation, which includes Refs. \cite{Taminiau2012,Abobeih2019,Bradley2019}.} for the proof). 
 
 The channels we consider in this Letter depend on certain parameters, and are diagonalizable (thus pseudo-Hermitian) for most parameter ranges. The diagonalizability of the channels [Eq. \eqref{Eq:Phi}] {will enable us} to well distinguish different eigenvalues in channel spectrum measurement. However, the channels may become non-diagonalizable at some isolated exceptional points (EPs), where the measurement statistics can display some transition, as discussed in the next section.
 
 %{In repetitive quantum channels, the diagonalizability of the channel enables the spectral decomposition [Eq. \eqref{Eq:Phi}] to distinguish different channel eigenvalues. For most cases we encounter below, the channel is diagonalizable, only with exceptions in some EPs (see \cite{Note1} for details). When the channel is influenced by certain parameters, such as coupling strength, dissipation rate, or evolution time, its spectrum typically varies accordingly. EPs of the quantum channel may thus emerge within critical regimes of these parameters. Since the quantum channel is pseudo-Hermitian, the EP describes a point in parameter space where a pair of complex conjugate eigenvalues coalesces into real eigenvalues. Then the effect can be noticed in the measurement statistics.}
 
 %{In the context we are considering, where we repeatedly operating the same quantum channel, the diagonalizability enables modal decomposition to distinguish different channel eigenvalues. For most cases we encounter, the channel is diagonalizable, with exceptions in some EPs (see \cite{Note1} for details).}

\nocite{Taminiau2012,Abobeih2019,Bradley2019}
 %One can also easily verify $\eta$ is a Hermitian operator, i.e., $\eta=\eta^\dagger$.

%We start from Eq. \eqref{Eq:Kraus}, when moving into HS space, it becomes $ \phii=\sum_{i=1}^r\mm_i$ with $\mm_i=M_i\otimes M_i^*$.  

%The quantum channel spectrum can exhibit abundant information including the parameters in the target system and the noise properties of a target gate \cite{Gu2023}.
%A quantum channel can be represented in terms of a set of Kraus operators $\{M_i\}_{i=1}^r$. 

\textit{Spectrum measurement of quantum channels.} We propose a general method to measure the spectrum of a quantum channel by repetitively tracking the measurement statistics [Fig.~\ref{Fig1}\blue{(a)}]. {The channel $\phii$ can be decomposed into a set of superoperators $\{\mm_i\}$. For $\mm_i=M_i\otimes M_i^*$ corresponding to a measurement outcome $i$, the system state collapses to $\mm_i\kett{\rho}/p_i$, where $p_i=\braa{\mathbb{I}}\mm_i\kett{\rho}={\rm Tr}(M_i\rho M_i^\dagger)$ is the probability to obtain such an outcome. }

We consider the evolution of probability to obtain outcome $i$ under repetitive quantum channels. The system state after $m$ channels becomes $\phii^m\kett{\rho}$, then the probability to get outcome $i$ in the $(m+1)$th cycle is 
\begin{equation}\label{pk}
\begin{aligned}
    p_i^{(m+1)}=\braa{\mathbb{I}}\mm_i \phii^{m}\kett{\rho}=\sum_{j=1}^{d^2}c_j\lambda_j^{m},
	 %p_i^{k+1}=\braa{\mathbb{I}}\mm_1 \sum_{j=1}^{d^2} \lambda_j^m\kett{R_j}\brakett{L_j}{\rho^1}=\sum_{j=1}^{d^2}c_j\lambda_j^m,
\end{aligned}
\end{equation}
with $c_j=\braa{\mathbb{I}}\mm_i \kett{R_j}\brakett{L_j}{\rho}={\rm Tr}(M_iR_jM_i^{\dagger}){\rm Tr}(L_j^{\dagger}\rho)$. If $\lambda_l=\lambda_k^*$, we have $c_l=\braa{\mathbb{I}}\mm_i |R_k^{\dagger}\rangle\rangle\langle\langle{L_k^{\dagger}}|{\rho}\rangle\rangle=c_k^*$. So the probability to {obtain} any outcome under repetitive channels can be expressed by {a polynomial function of the channel spectrum} [Fig.~\ref{Fig1} \blue{(b)}]. 
Note that a real eigenvalue $\lambda_j$ contributes an exponential decay $c_j\lambda_j^m$ {as a function of the repetition number $m$}, while a pair of complex conjugate eigenvalues $\{\lambda_k,\lambda_k^*\}$ induce a damped oscillation $2{\rm Re}(c_k)|\lambda_k|^m\cos(m\varphi_k)$ with $|\lambda_k|<1$ and $\varphi_k=\arg{\lambda_k}$. {Moreover, the initial state should be carefully chosen so that more elements in $\{c_j\}$ are nonzero, for example, $c_j\neq0$ requires ${\rm Tr}(L_j^{\dagger}\rho)\neq0$.}
%{Similar relationship can also be observed in the statistics of two-point measurement correlation function, as illustrated in the Supplementary Materials \cite{Note1}.}

\begin{figure*}[htbp]
\centering
  \includegraphics[width=17cm]{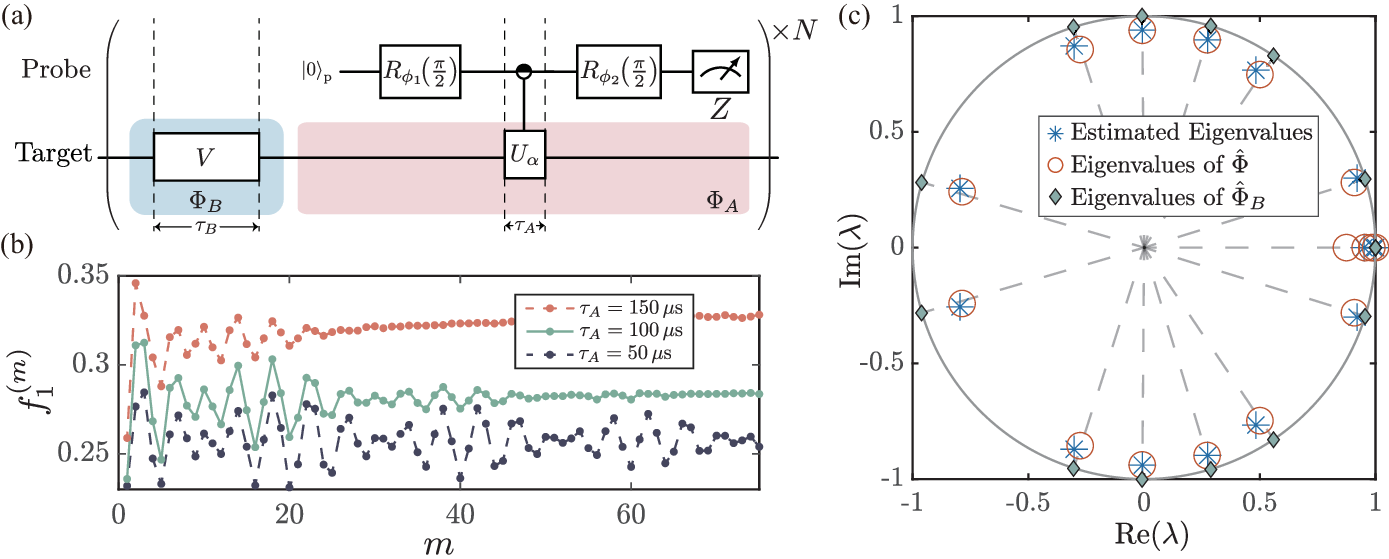}
  \caption{
Applying quantum channel spectrum measurement to Hamiltonian parameter estimation. (a) Illustration of the quantum circuit for Hamiltonian parameter estimation. The target system evolves under repetitive quantum channels, with each channel $\phii$ concatenated by $\phii_B$ generated by a free evolution and $\phii_A$ induced by a probe qubit under RIM. 
  (b) Frequency $f_1^{(m)}$ as a function of measurement number $m$ with different RIM evolution time $\tau_A$. The oscillation damping rate increases with $\tau_A$ (corresponding to increasing measurement strength).
  %{\color{red} (add another one or two oscilations with different $\tau_A$, and show a single one in c)} 
  (c) Comparison of the ideal spectra of $\phii_B$ (blue diamonds), $\phii$ (red circles) and the estimated spectra of $\phii$ (blue stars). Under the perturbation of $\phii_A$, the eigenvalues of $\phii$ have almost the same phase angles as those of $\phii_B$ but reduced amplitudes. Parameters are $|\vb*{h}_1|/2\pi=1.20$ kHz, $|\vb*{h}_2|/2\pi=1.33$ kHz, $D/2\pi=105.34$ Hz, $\omega/2\pi=1$ kHz, $\tau_A=100\,\,\mu$s and $\tau_B=227.3\,\,\mu$s.
  %(b) The reconstructed EP line. The EPs labeled by diamond are gained by scanning $\nu$ with certain $\mu$, which is shown in the inset with $\mu=0.25\pi$. While those labeled by circles are gained conversely. The solid lines indicate the theoretical results. The data are obtained by Monte Carlo simulations with $10^5$ samples. 
  %The estimated eigenvalues obtained by matrix pencil methods are labeled as blue stars.
 }\label{NV}
\end{figure*}

As $N$ repetitive quantum channels can be decomposed as $r^N$ stochastic quantum trajectories \cite{Ma2023g,Jin2024,Qiu2024,Zhang2024}, i.e., $ \phii^N=\sum_{i_1,i_2 \ldots, i_N=0}^r \hat{\mathcal{M}}_{i_N} \cdots \hat{\mathcal{M}}_{i_2}\hat{\mathcal{M}}_{i_1}$, we can sample a sufficient number of trajectories such that $\{p_i^{(m)}\}_{m=1}^N$ can be well approximated by $\{f_i^{(m)}\}_{m=1}^N$ with $f_i^{(m)}$ being the frequency of outcome $i$ in the $m$th measurement among these trajectories. {According to the Hoeffding's inequality \cite{hoeffding1963probability}, the number of sampled trajectories} should be at least $N_s=\frac{1}{2\delta^2}\ln(\frac{2}{\epsilon}) $ to estimate each point within a fixed accuracy $\delta$ with probability $1-\epsilon$. Then $\{\lambda_j\}$ can be accurately extracted by the matrix pencil (MP) method \cite{Sarkar1995} [Fig.~\ref{Fig1}\blue{(c)}].
{We note that the channel spectrum can also be characterized by complete channel tomography (or quantum process tomography) \cite{Kunjummen2023}, which requires complete direct control of the quantum system, including the system preparation in a complete set of initial states and the ability to perform quantum state tomography after the channel. In contrast, our method for channel spectrum measurement applies when we have only limited indirect control on the quantum system assisted by an ancilla system, as demonstrated below.}

Moreover, EPs of the quantum channel can manifest themselves in the measurement statistics of repetitive channels. For a pseudo-Hermitian quantum channel, EPs are the points in parameter space where a pair of complex conjugate eigenvalues coalesces into real eigenvalues.  {Then near the EPs, the measurement statistics of $\{p_i^{(m)}\}$ as a function of $m$ can exhibit the crossover from damped oscillations (from eigenvalues in complex conjugate pairs) to exponential decays (from real eigenvalues), as shown in Sec. S2 of SM \cite{Note1}.}

%\textit{Detecting the free Hamiltonian of target system by a Ramsey channel (?).} 
\textit{Application to Hamiltonian parameter estimation.} 
{As an application, we show that the spectrum measurement of quantum channels can be utilized for learning the parameters in the free Hamiltonian of a target system with a known eigenstructure.} We consider a typical class of quantum channels, which is the concatenation of a unitary channel $\Phi_B$ generated by the free Hamiltonian of the target system and a channel $\Phi_A$ induced by a Ramsey interferometry measurement (RIM) of a probe [Fig.~\ref{NV} \blue{(a)}]. 

The unitary channel on the target system is induced by a free Hamiltonian $B=\sum_{i}b_i|i\rangle\langle i|$ with a free evolution time $\tau_B$, i.e., $\Phi_B(\rho)=V\rho V^\dagger$ with $V=\exp{-\ii B\tau_B}$. In the vectorized operator space, we have $\phii_B=\sum_{ij}v_{ij}\kett{ij}\braa{ij}$ with $v_{ij}=\ee^{-\ii\bij\tau_B}$ and $\bij=b_i-b_j$.
%The RIM sequence is commonly used to measure the coherence of a quantum system, realize high-resolution single-spin NMR and track the spin precession \cite{cujia2019,pfender2019}. 
The weak-measurement channel on the target system is induced by the commonly-used RIM sequence. In each RIM, a probe qubit is first initialized to $\ket{0}_\q$, and then rotated to $\ket{\psi}_\q=R_{\phi_1}(\frac{\pi}{2})\ket{0}_\q$, where $R_{\phi}(\theta)=\ee^{-\ii(\cos\phi \sigma_\q^x+\sin\phi\sigma_\q^y)\theta/2}$ is the rotation operator, $\sigma_\q^i\,(i=x,y,z)$ is the Pauli-$i$ matrix of the probe qubit and $\sigma_\q^z=|0\rangle_\q\langle0|-|1\rangle_\q\langle1|$. After that, the probe interacts with a target system with the coupling Hamiltonian $H_A=\sigma_\q^z\otimes A$ for time $\tau_A$. 
{Then} the probe is rotated by $R_{\phi_2}(\frac{\pi}{2})$ before a {final} projective measurement in the basis of $\sigma_\q^z$. For the probe measurement outcome $\alpha\in\{0,1\}$, the target undergoes the operation $\mm_\alpha$ with $M_{\alpha}=[U_0-(-1)^{\alpha} \ee^{\ii\phi}U_1]/2$, $U_{\alpha}=\exp{-\ii(-1)^\alpha A \tau_A }$ {and $\phi=\phi_1-\phi_2$}. The channel induced by the RIM on the target system is $\phii_{A}=\sum_{\alpha=0}^1 \mm_\alpha$ \cite{Ma2023g,Jin2024,Qiu2024,Zhang2024}. {Here we assume that the probe rotations take much shorter time than $\tau_A$ and the free Hamiltonian is not included during the RIM. In Sec. S3 of SM \cite{Note1}, we show the effect of the free Hamiltonian during the RIM can be separated out in the weak-measurement regime to contribute additional free evolution of the target system. Alternatively, the effective free Hamiltonian can be averaged out with additional dynamical decoupling (DD) control of the probe \cite{pfender2019,cujia2019}.}

%, with $M_\alpha=[U_0-(-1)^{\alpha} e^{\ii\phi}U_1]/2$ and $U_{\alpha}=\exp{-i(-1)^\alpha A \tau_A }$. 

%The quantum Zeno effect can also be incorporated into this "evolution-measurement" model. Zeno dynamics would arise  which can project it onto the eigenspaces.. However, here we mainly consider a weak entanglement between the probe and the target system to track the evolution of the target.

%To track the free evolution of $\phii_B$, $\phii_A$ should constitute a weak measurement on the system, which can be realized if $\tau_A$ is rather short.

%and giving $\hat{\mathcal C}_A|Y\rangle\rangle=|[A,Y]\rangle\rangle$

Quantum Zeno dynamics can arise if the RIMs {induce} strong and frequent measurements on the target system \cite{fischer2001observation,chaudhry2016general,li2013quantum}. However, here we use repetitive weak measurements to track the evolution $\phii_B$ of the target system. $\phii_A$ constitutes a weak measurement on the target if {$\tau_A||A||\ll1$ with $||A||=\max{\{\sqrt{\langle\psi|A^{\dagger}A|\psi\rangle}/\sqrt{\langle \psi|\psi\rangle}: |\psi\rangle\neq 0\}}$ denoting the spectral norm.} Then $\phii_A$ can be perturbatively expanded up to the second order of $\tau_A$ as $\phii_A\approx \mathbb{I}\otimes \mathbb{I}+\tau_A^2\hat{\mathcal{L}}$,
%\begin{equation}
  %\phii_A= I\otimes I-\frac{\tau_A^2}{2}\hat K,
%  \phii_A= I\otimes I+\tau_A^2\hat L,
%\end{equation}
where $\hat{\mathcal{L}} = A\otimes A^T-\{A^2\otimes\mathbb{I} + \mathbb{I}\otimes (A^T)^2\}/2 $ is a Lindbladian with $A$ being the jump operator. Note that $\hat{\mathcal{L}}= -\hat{\mathcal C}_A^2/2$ with $\hat{\mathcal C}_A=A\otimes \mathbb{I}-\mathbb{I}\otimes A^T$ being the vectorization of the commutator $[A,\cdot]=A(\cdot) \mathbb{I}-\mathbb{I}(\cdot) A$. 
Then the concatenated channel is 
\begin{equation}
  \phii=\phii_A\phii_B\approx\phii_B+\tau_A^2\hat{\mathcal{L}}\phii_B,
\end{equation}
where the second term can be regarded as a perturbation on $\phii_B$. The eigenvalues $\{\lambda_{ij}\}$ ($i\neq j$) of $\phii$ up to the first-order perturbation are
%where $\hat L = A^2\otimes I + I\otimes (A^T)^2-2 A\otimes A^T= \hat{\mathcal C}_A^2$ is a superoperator in the HS space and $\hat{\mathcal C}_A=A\otimes I-I\otimes A^T$ is the vectorization of the commutator $[A,\cdot]=A(\cdot) I-I(\cdot) A$, giving $\hat{\mathcal C}_A|Y\rangle\rangle=|[A,Y]\rangle\rangle$. We note that $-\hat K$ is also a Lindbladian, with the Hermitian operator $A$ being the Lindblad dissipator. When we consider $\phii_A$ as a perturbation on $\phii_B$, we can expand the channel in the basis of $B$ ($B=\sum_{k} b_{k}\ket{k}\bra{k}$), then we have $\phii_B=\sum_{ij}V_{ij}\kett{ij}\braa{ij}$ with $V_{ij}=\ee^{-\ii(b_i-b_j)\tau_B}$.  According to the perturbation theory, the first order perturbation of the eigenvalues of $\phii$ can be obtained as
\begin{equation}\label{Vij}
  \lambda_{ij}\approx v_{ij}\left(1-\frac{\tau_A^2}{2}\braa{ij}\hat{\mathcal C}_A^2\kett{ij}\right).
\end{equation}
Since $A$ and $\hat{\mathcal C}_A$ are both Hermitian, $\braa{ij}\hat{\mathcal C}_A^2\kett{ij}$ is a non-negative real number. Then compared to the eigenvalue $v_{ij}$ of $\phii_B$, the eigenvalue $\lambda_{ij}$ of $\phii$ has the same phase angle but a reduced amplitude, which results in a damped oscillation in Eq. \eqref{pk}. Such a property has enabled error mitigation for quantum phase estimation algorithms \cite{Gu2023a}. Since $\phii$ contains two {superoperators} $\{\mm_0\phii_B,\mm_1\phii_B\}$ corresponding to the two probe measurement outcomes, the spectrum of $\phii_B$ or $B$ can be well estimated by tracking the measurement statistics (e.g., $f_1^{(m)}$) of the probe. {For the signal $f_1^{(m)}\approx \sum_{i,j=1}^{d} c_{ij}\lambda_{ij}^m$, the amplitude $c_{ij}$ with $i\neq j$ grows linearly with $\tau_A$, so $\tau_A$ cannot be too small. For $B$ with a known form (or eigenstructure), we can thus estimate the unknown parameters in $B$. All the details about the perturbation theory of quantum channels and amplitudes of the weak- measurement signals can be found in Sec. S3 of SM \cite{Note1}. }

%{Specifically, our protocol estimates the frequency differences $v_{ij}$ from the channel spectrum. Parameter recovery can be realized whenever the mapping form between the parameters and the eigenvalues can be fully determined. This usually requires the full knowledge of the eigenstructure.}

%We also note that $\Phi(\cdot)=U_B(\Phi_A(\cdot))U_B^\dagger=\sum_{\alpha=0}^1 \tilde{M}_\alpha\rho \tilde{M}_\alpha^\dagger$, in which $\tilde{M}_\alpha=U_B{M}_\alpha$, and thus $\tilde p_i=\Tr(U_B{M}_\alpha\rho {M}^\dagger_\alpha U^\dagger_B)=p_i$. However, $\tilde p_1^{m+1}=\braa{\mathbb{I}}\tilde\mm_1\phii^m\kett{\rho^1}=\braa{\mathbb{I}}\mm_1 \sum_{j=1}^{d^2} \lambda_j^m\kett{R_j}\brakett{L_j}{\rho^1}=\sum_{j=1}^{d^2}c_j\lambda_j^m,$Hence the results of RIMs can give the spectra of the total channel.

\textit{Example I: A probe spin coupled to a target spin.}
We first demonstrate that our method can be directly used to sense the precession frequency of a target spin. The target system first evolves {for time $\tau_B$} under the free Hamiltonian $B=\omega\sigma_z/2$, with $\omega$ being the Larmor frequency. Then during each RIM, the probe qubit is coupled to the target spin {for time $\tau_A$} with the Hamiltonian $H_A=g\sigma^z_\q\otimes\sigma_x/2$, where $g$ is the coupling strength. %{ We note that the free evolution in this model can be turned off by employing dynamical decoupling sequences in the RIM period \cite{cujia2019,pfender2019}.}
So the concatenated channel $\phii$ is
\begin{equation}
  \hat \Phi=
\mqty(
  \cos^2\left(\frac{\mu}{2}\right) & 0 & 0 & \sin^2\left(\frac{\mu}{2}\right) \\
  0 & \ee^{-\ii\nu} \cos^2\left(\frac{\mu}{2}\right) & \ee^{\ii\nu} \sin^2\left(\frac{\mu}{2}\right) & 0 \\
  0 & \ee^{-\ii\nu} \sin^2\left(\frac{\mu}{2}\right) & \ee^{\ii\nu} \cos^2\left(\frac{\mu}{2}\right) & 0 \\
  \sin^2\left(\frac{\mu}{2}\right) & 0 & 0 & \cos^2\left(\frac{\mu}{2}\right)
)
\end{equation}
with $\mu=g\tau_A$ and $\nu=\omega \tau_B$. We can analytically obtain the channel spectrum, containing a fixed point $\lambda_1=1$, three decaying points $\lambda_2=\cos(\mu)$ and $\lambda_{3,4}=\cos^2\left(\frac{\mu}{2}\right)\left[\cos\nu\pm\sqrt{\tan^4\left(\frac{\mu}{2}\right)-\sin^2\nu}\right]$. For small $\mu$, the expansion of $\lambda_{3,4}$ agrees with Eq. \eqref{Vij}, then the Larmor frequency $\omega$ can be detected by channel spectrum measurement. For arbitrary $\mu$ and $\nu$, an EP line lies on the line $\tan^4\left(\frac{\mu}{2}\right)=\sin^2\nu$ in the $(\mu,\nu)$ plane of the parameter space.
%\nocite{Taminiau2012,Abobeih2019,Bradley2019}
%By scanning the parameter space and estimating the corresponding eigenvalues, we can also locate EPs in the system and construct the phase diagram \cite{Note1}.
{We note that a similar example has been considered in Refs. \cite{cujia2019,pfender2019}, however, without realizing the connection with channel spectrum measurement and the effects of EPs on the measurement statistics. }

%{While previous works focused on this fully analytical model, the theoretical framework in this letter can be used to measure the spectrum of any channel and thus sense the parameters in the Hamiltonian of an arbitrary target system.} %The theoretical insights in this paper enables us to sense a generic Hamiltonian of an arbitrary quantum system.
%Given a known Hamiltonian, one can further obtain the parameters in the Hamiltonian, without the knowledge of the whole quantum channel.

\textit{Example II: A probe spin coupled to a spin cluster.}
We then consider a central probe spin coupled to a nuclear spin cluster containing $M$ nuclear spins. With a strong magnetic field, the Hamiltonians are $A=\sum_{k=1}^M \vb*{h}_k\cdot \vb*{I}_k$ and $B\approx\omega\sum_{k=1}^M I^z_k+\sum_{j<k}D_{jk}(I_j^+I_k^-+I_j^-I_k^+-4I_j^z I_k^z)$, where $\vb*{h}_k=(h_k^x, h_k^y, h_k^z)$ is hyperfine coupling parameter, $D_{jk}$ denotes the dipolar coupling strength, $\vb*{I}_k=(I_k^x, I_k^y, I_k^z)$ is the $k$th nuclear spin-1/2 operator and $I_k^{\pm}=I^x_k\pm \ii I^y_k$. Then for $M=2$, the spectra of $B$ and $\phii_B$ are $\{b_i\}=\{0,-D\pm \omega,2D\}$ and $\{\bij\}=\{\pm 2D,\pm (\omega-3D),\pm (\omega-D),\pm (\omega+D),\pm (\omega+3D),\pm 2\omega,0\}$ with $D=D_{12}$. In the weak measurement regime with $|\vb*{h}_k|\tau_A\ll 1$, the parameters $\omega$ and $D$ can be estimated by measuring the spectrum of $\phii$. {We perform simulations with the parameters chosen as $\omega/2\pi=1$ kHz and $D/2\pi=105.34$ Hz.} The results in Fig.~\ref{NV}\blue{(b)} show the average signal of $f_1$ over $10^6$ samples of quantum trajectories, which is composed of multiple modes of damped oscillations. For this type of signal, we can use the MP method to accurately extract the oscillation frequency and decaying rate of each mode [Fig. \ref{NV}\blue{(c)}]. We can also see that the phase angles of eigenvalues are almost not perturbed, so that the parameters $\{\bij\}$ can be accurately inferred from the estimated spectrum $\{\tilde\lambda_{ij}\}$ of $\phii_B$. In Table~\ref{Table}, we list the estimated phases $\tilde\phi_{ij}=\ln(\tilde \lambda_{ij}/|\tilde \lambda_{ij}|)$, the estimated parameters $\tilde \bij$ as $\tilde\phi_{ij}/\tau_B$, and the corresponding Hamiltonian parameters. 

\begin{table}[htbp]
\caption{Estimated phases and the corresponding Hamiltonian parameters for a two-spin cluster. The estimated value of parameters are $\tilde \omega/2\pi=1003.2$ Hz, $\tilde D/2\pi=106.19$ Hz with the estimation errors {between the estimated parameters and the actual parameters} being 0.3\% and 0.8\%, respectively.
%the relevant errors are $\delta(\omega)=0.3\%$ and $\delta(D)=0.8\%$
}
\begin{ruledtabular}
\begin{tabular}{lll}
Phases ($^\circ$) & $\tilde \bij/2\pi$ (kHz)  &  Parameters \\
\hline
18.07 & 220.90 & $2D$\\
57.79 & 706.31 & $\omega-3D$\\
72.97 & 891.88 & $\omega-D$\\
90.46 & 1105.6 & $\omega+D$\\
109.1 & 1333.6 & $\omega +3D$\\
162.2 & 1982.2 & $2\omega$\\
\end{tabular}
\end{ruledtabular}\label{Table}
\end{table}

%We obtain the estimated value of parameters $\omega/2\pi=1003.2$ Hz, $D/2\pi=106.19$ Hz, and the relevant errors are $\delta(\omega)=0.3\%$ and $\delta(D)=0.8\%$.

%We list the estimated phases, which is $\tilde\phi_i=\ln(\tilde \lambda_i/|\tilde \lambda_i|)$ with estimated eigenvalues $\tilde \lambda_i$, $\tilde \bij/2\pi=\tilde\phi_i/(2\pi\tau_B)$, and corresponding quantities in Table~\ref{Table}.  Through the table, we obtain the estimated value of parameters $\omega/2\pi=1003.2$ Hz, $D/2\pi=106.19$ Hz, and the relevant errors are $\delta(\omega)=0.3\%$ and $\delta(D)=0.8\%$.

% \begin{figure*}[htbp]
% \centering
%   \includegraphics[width=16cm]{Figure3.pdf}
%   \caption{Quantum channel eigenvalues estimation as a tool to obtain Hamiltonian parameters. The (a) original fluctuation data of $f_1$ and (b) corresponding Fourier transformed power spectrum with the evolution of measurements time $m$. (c) The eigenvalues of $\phii_B$ (purple dots) are located on the unit circle (grey circle) in the complex plane. Under the perturbation of $\phii_A$, the absolute values of the eigenvalues of $\phii$ (red circles) relax to the decaying points. The estimated eigenvalues obtained by matrix pencil methods are labeled as blue stars. Parameters are $A_1/2\pi=1.20$ kHz, $A_2/2\pi=1.33$ kHz, $D/2\pi=105.34$ Hz, $\omega/2\pi=1$ kHz, $\tau_A=100\,\,\mu$s and $\tau_B=227.3\,\,\mu$s.}
%   \label{Fig:Simulation}
% \end{figure*}

\textit{Experimental considerations.}  %\color{red} Discuss the relation between the model in our paper with previous experimental works. Show that our method can be used for nuclear magnetic resonance (NMR) of nuclear spin clusters, which may help realize single-molecule NMR. Also discuss the advantages of such methods, as discussed in previous experiments.
Finally we discuss the feasibility of our spectrum measurement scheme for nanoscale NMR with solid-state defect systems, such as the nitrogen-vacancy (NV) center system in diamond. Experiments with models similar to Example I have been reported in \cite{cujia2019,pfender2019}, where an NV electron spin under dynamical decoupling sequences repetitively tracks the precession of a single nuclear spin in diamond. Compared to conventional dynamical decoupling spectroscopy whose resolution is mainly limited by the probe coherence time \cite{Lange2011,Mamin2013,Loretz2013}, these schemes can achieve higher spectral resolution due to the much longer coherence time of the target system. Moreover, the weak measurements induced by the probe do not perturb the estimated frequencies, so these schemes also have much higher accuracy.
However, the theoretical models in these works only apply to a single spin-1/2 target, and cannot be extended to more complex nuclear spin clusters. The theoretical framework in this Letter fills this gap, so that we can accurately sense the Hamiltonian parameters of a complex nuclear-spin cluster as in Example II. 

%{In the Supplementary Material \cite{Note1}, we also numerically verify the scheme is feasible with the presence of target system noise, and can accurately detect a nuclear spin cluster containing more spins.}
%{The scheme is also feasible with the presence of Lindblad noises, e.g., probe and target dephasing and relaxation, since this type of noises don't adjust the phases of the eigenvalues. However, additional damping would cause the further decaying of the signal amplitude.}

%Unlike  the spectral resolution is not  $T_{2,\text{DD}}$ since the evolution time $\tau_A$ is small in the sequential weak measurement scheme.
In practical experiments with a spin-1 NV probe electron spin with the basis states $\{\ket{0}_\mathrm{e},\ket{\pm 1}_\mathrm{e}\}$, {the initial state of the target system should have some polarization (purity), i.e., cannot be the maximally mixed state, which can be realized by sequential quantum non-demolition measurements aided by the probe spin.} For the weak-measurement channel induced by an RIM sequence, we can use the subspace $\{\ket{+1}_\mathrm{e}, \ket{-1}_\mathrm{e}\}$ of the probe, while the probe is initialized to state $\ket{0}_\mathrm{e}$ to avoid interaction with the target system during the free-evolution channel. 

{In Secs. S3 and S4 of SM \cite{Note1}, we perform extensive analyses and simulations to assess the practical feasibility of this scheme. We first analyze the effect of residual free Hamiltonian during the measurement channel. As the hyperfine couplings between the probe spin and nuclear spins are often much stronger than the dipole-dipole couplings between nuclear spins, we prove that such a residual Hamiltonian has negligible effects on the spectrum measurement. Then we numerically verify the scheme is feasible with the presence of target system noise, and can accurately detect a nuclear spin cluster containing more spins. Moreover, we discuss that the dynamical decoupling method can also be incorporated into this scheme to selectively sense a subset of coupling parameters in a large spin cluster. 
}

%Furthermore, when extracting the Hamiltonian parameters, the scheme can resolve very low coupling parameters (at about $300$ Hz) for the target system even with strong coupling between the probe and the target (at about $30$ kHz) \cite{Note1}.

\textit{Conclusions and outlooks.} {We have presented a general scheme to measure the channel spectrum by tracking the measurement statistics of repetitive quantum channels. Based on perturbation theory, we analyze a typical class of concatenated quantum channels, with each channel consisting of a unitary channel and a weak-measurement channel, and thus provide a general framework for previous experimental works. Then we demonstrate that such a scheme can perform Hamiltonian parameter estimation, which is potentially useful for nanoscale and even single-molecule NMR \cite{Du2024}. 
The scheme can be extended to measure the spectra of other quantum channels induced by more general couplings between the quantum system and its environment, which may be potentially useful for efficient quantum system learning \cite{Gebhart2023}. It will also be interesting to study the role of symmetry and topology \cite{Gong2018,Nakagawa2025} of  quantum channels and explore the applications of EPs in this scheme.}

%\color{red} Disucss that our work uncovers the interesting non-Hermitian physics in quantum channels. Such pseudo-Hermitian quantum channels can be easily constructed physically and can have applications in quantum system learning. Related open problems include the conserved quantities of quantum channels due to the pseudo-Hermiticity,  the physics of exceptional points, the physics of quantum channels induced by general couplings between an ancilla and a target system,etc. 
%We have uncovered the NH physics in quantum channels and its potential applications. 
%We prove that a generic quantum channel is pseudo-Hermitian if it is diagonalizable.  %Related open problems include the symmetry and topology \cite{Gong2018,Nakagawa2025} of pseudo-Hermitian quantum channels and the understandings and applications of EP points in such channels.
%This finding may provide a new playground for the fundamental areas of PT-symmetric and pseudo-Hermitian quantum mechanics, as quantum channels can describe a much broader range of practical scenarios than NH Hamiltonians and Lindbladians.
%Note that this scheme applies to any quantum channel that is not necessarily pseudo-Hermitian. 

We thank Ryusuke Hamazaki for providing very helpful comments. The research is supported by the National Natural Science Foundation of China (No. 12574082, No. 12174379, No. E31Q02BG), the Chinese Academy of Sciences (No. E0SEBB11, No. E27RBB11), Quantum Science and Technology-National Science and Technology Major Project (No. 2021ZD0302300) and Chinese Academy of Sciences Project for Young Scientists in Basic Research (YSBR-090).

\textit{Data availability.} The data that support the findings of this article are openly available \footnote{Y.-D. Jin and W.-L. Ma, Data for paper ``Spectrum measurement of quantum channels and application to Hamiltonian parameter estimation'', Zenodo, 2025, \url{https://doi.org/10.5281/zenodo.17595678}}.

\bibliography{NHP.bib}

@book{Nielsen2010,
  title={Quantum computation and quantum information},
  author={Nielsen, Michael A and Chuang, Isaac L},
  year={2010},
  publisher={Cambridge university press}
}

@book{rivas2012,
  title={Open quantum systems},
  author={Rivas, Angel and Huelga, Susana F},
  volume={10},
  year={2012},
  publisher={Springer}
}

@article{Kunjummen2023,
  title = {Shadow process tomography of quantum channels},
  author = {Kunjummen, Jonathan and Tran, Minh C. and Carney, Daniel and Taylor, Jacob M.},
  journal = {Phys. Rev. A},
  volume = {107},
  issue = {4},
  pages = {042403},
  numpages = {12},
  year = {2023},
  month = {Apr},
  publisher = {American Physical Society},
  doi = {10.1103/PhysRevA.107.042403},
  url = {https://link.aps.org/doi/10.1103/PhysRevA.107.042403}
}

@article{Gong2018,
  title = {Topological Phases of Non-Hermitian Systems},
  author = {Gong, Zongping and Ashida, Yuto and Kawabata, Kohei and Takasan, Kazuaki and Higashikawa, Sho and Ueda, Masahito},
  journal = {Phys. Rev. X},
  volume = {8},
  issue = {3},
  pages = {031079},
  numpages = {33},
  year = {2018},
  month = {Sep},
  publisher = {American Physical Society},
  doi = {10.1103/PhysRevX.8.031079},
  url = {https://link.aps.org/doi/10.1103/PhysRevX.8.031079}
}

@article{Malzard2015,
  title = {Topologically Protected Defect States in Open Photonic Systems with Non-Hermitian Charge-Conjugation and Parity-Time Symmetry},
  author = {Malzard, Simon and Poli, Charles and Schomerus, Henning},
  journal = {Phys. Rev. Lett.},
  volume = {115},
  issue = {20},
  pages = {200402},
  numpages = {6},
  year = {2015},
  month = {Nov},
  publisher = {American Physical Society},
  doi = {10.1103/PhysRevLett.115.200402},
  url = {https://link.aps.org/doi/10.1103/PhysRevLett.115.200402}
}

@article{Rivero2020,
  title = {Pseudochirality: A Manifestation of Noether's Theorem in Non-Hermitian Systems},
  author = {Rivero, Jose D. H. and Ge, Li},
  journal = {Phys. Rev. Lett.},
  volume = {125},
  issue = {8},
  pages = {083902},
  numpages = {6},
  year = {2020},
  month = {Aug},
  publisher = {American Physical Society},
  doi = {10.1103/PhysRevLett.125.083902},
  url = {https://link.aps.org/doi/10.1103/PhysRevLett.125.083902}
}

@article{Garc2022,
  title = {Symmetry Classification and Universality in Non-Hermitian Many-Body Quantum Chaos by the Sachdev-Ye-Kitaev Model},
  author = {Garc\'{\i}a-Garc\'{\i}a, Antonio M. and S\'a, Lucas and Verbaarschot, Jacobus J. M.},
  journal = {Phys. Rev. X},
  volume = {12},
  issue = {2},
  pages = {021040},
  numpages = {22},
  year = {2022},
  month = {May},
  publisher = {American Physical Society},
  doi = {10.1103/PhysRevX.12.021040},
  url = {https://link.aps.org/doi/10.1103/PhysRevX.12.021040}
}

@article{Kawabata2019,
  title = {Symmetry and Topology in Non-Hermitian Physics},
  author = {Kawabata, Kohei and Shiozaki, Ken and Ueda, Masahito and Sato, Masatoshi},
  journal = {Phys. Rev. X},
  volume = {9},
  issue = {4},
  pages = {041015},
  numpages = {52},
  year = {2019},
  month = {Oct},
  publisher = {American Physical Society},
  doi = {10.1103/PhysRevX.9.041015},
  url = {https://link.aps.org/doi/10.1103/PhysRevX.9.041015}
}

@article{Halder2024,
doi = {10.1088/1361-648X/ad4940},
url = {https://dx.doi.org/10.1088/1361-648X/ad4940},
year = {2024},
month = {may},
publisher = {IOP Publishing},
volume = {36},
number = {33},
pages = {335301},
author = {Halder, Dipendu and Basu, Saurabh},
title = {Parsing skin effect in a non-Hermitian spinless BHZ-like model},
journal = {Journal of Physics: Condensed Matter}
}

@article{Du2024,
  title = {Single-molecule scale magnetic resonance spectroscopy using quantum diamond sensors},
  author = {Du, Jiangfeng and Shi, Fazhan and Kong, Xi and Jelezko, Fedor and Wrachtrup, J\"org},
  journal = {Rev. Mod. Phys.},
  volume = {96},
  issue = {2},
  pages = {025001},
  numpages = {62},
  year = {2024},
  month = {May},
  publisher = {American Physical Society},
  doi = {10.1103/RevModPhys.96.025001},
  url = {https://link.aps.org/doi/10.1103/RevModPhys.96.025001}
}

@article{Albert2014,
  title = {Symmetries and Conserved Quantities in {{Lindblad}} Master Equations},
  author = {Albert, Victor V. and Jiang, Liang},
  year = {2014},
  month = feb,
  journal = {Phys. Rev. A},
  volume = {89},
  number = {2},
  pages = {022118},
  publisher = {American Physical Society},
  doi = {10.1103/PhysRevA.89.022118}
}

@article{Gu2023,
  title={Benchmarking universal quantum gates via channel spectrum},
  author={Gu, Yanwu and Zhuang, Wei-Feng and Chai, Xudan and Liu, Dong E},
  journal={Nature Communications},
  volume={14},
  number={1},
  pages={5880},
  year={2023},
  publisher={Nature Publishing Group UK London}
}

@article{Gu2023a,
  title = {Noise-Resilient Phase Estimation with Randomized Compiling},
  author = {Gu, Yanwu and Ma, Yunheng and Forcellini, Nicol\`o and Liu, Dong E.},
  journal = {Phys. Rev. Lett.},
  volume = {130},
  issue = {25},
  pages = {250601},
  numpages = {6},
  year = {2023},
  month = {Jun},
  publisher = {American Physical Society},
  doi = {10.1103/PhysRevLett.130.250601},
  url = {https://link.aps.org/doi/10.1103/PhysRevLett.130.250601}
}

@article{ashidaNonHermitianPhysics2020,
  title = {Non-{{Hermitian}} Physics},
  author = {Ashida, Yuto and Gong, Zongping and Ueda, Masahito},
  year = {2020},
  month = jul,
  journal = {Adv. Phys.},
  volume = {69},
  number = {3},
  pages = {249--435},
  issn = {0001-8732, 1460-6976},
  doi = {10.1080/00018732.2021.1876991},
  urldate = {2025-08-19},
  langid = {english}
}

@article{bergholtz2021,
  title = {Exceptional Topology of Non-{{Hermitian}} Systems},
  author = {Bergholtz, Emil J. and Budich, Jan Carl and Kunst, Flore K.},
  year = {2021},
  month = feb,
  journal = {Rev. Mod. Phys.},
  volume = {93},
  number = {1},
  pages = {015005},
  issn = {0034-6861, 1539-0756},
  doi = {10.1103/RevModPhys.93.015005},
  urldate = {2025-08-19},
  langid = {english}
}

@article{Bidhi2025,
  title = {Quantum-classical correspondence in quantum channels},
  author = {Vijaywargia, Bidhi and Lakshminarayan, Arul},
  journal = {Phys. Rev. E},
  volume = {111},
  issue = {1},
  pages = {014210},
  numpages = {24},
  year = {2025},
  month = {Jan},
  publisher = {American Physical Society},
  doi = {10.1103/PhysRevE.111.014210},
  url = {https://link.aps.org/doi/10.1103/PhysRevE.111.014210}
}

@article{blok2014manipulating,
  title = {Manipulating a Qubit through the Backaction of Sequential Partial Measurements and Real-Time Feedback},
  author = {Blok, {\relax MS} and Bonato, Cristian and Markham, {\relax ML} and Twitchen, {\relax DJ} and Dobrovitski, {\relax VV} and Hanson, R},
  year = {2014},
  journal = {Nat. Phys.},
  volume = {10},
  number = {3},
  pages = {189--193},
  publisher = {Nature Publishing Group UK London},
  doi = {10.1038/nphys2881}
}

@article{Breuer2016,
  title = {Colloquium: {{Non-Markovian}} Dynamics in Open Quantum Systems},
  author = {Breuer, Heinz-Peter and Laine, Elsi-Mari and Piilo, Jyrki and Vacchini, Bassano},
  year = {2016},
  month = apr,
  journal = {Rev. Mod. Phys.},
  volume = {88},
  number = {2},
  pages = {021002},
  publisher = {American Physical Society},
  doi = {10.1103/RevModPhys.88.021002}
}

@article{Caruso2014,
  title = {Quantum Channels and Memory Effects},
  author = {Caruso, Filippo and Giovannetti, Vittorio and Lupo, Cosmo and Mancini, Stefano},
  year = {2014},
  month = dec,
  journal = {Rev. Mod. Phys.},
  volume = {86},
  number = {4},
  pages = {1203--1259},
  publisher = {American Physical Society},
  doi = {10.1103/RevModPhys.86.1203}
}

@article{chaudhry2016general,
  title = {A General Framework for the Quantum {{Zeno}} and Anti-{{Zeno}} Effects},
  author = {Chaudhry, Adam Zaman},
  year = {2016},
  journal = {Sci. Rep.},
  volume = {6},
  number = {1},
  pages = {29497},
  publisher = {Nature Publishing Group UK London},
  doi = {10.1038/srep29497}
}

@article{Chen2025,
  title = {Engineering Nonequilibrium Steady States through Floquet Liouvillians},
  author = {Chen, Weijian and Abbasi, Maryam and Erdamar, Serra and Muldoon, Jacob and Joglekar, Yogesh N. and Murch, Kater W.},
  year = {2025},
  month = mar,
  journal = {Phys. Rev. Lett.},
  volume = {134},
  number = {9},
  pages = {090402},
  publisher = {American Physical Society},
  doi = {10.1103/PhysRevLett.134.090402}
}

@article{ciccarello2022,
  title = {Quantum Collision Models: {{Open}} System Dynamics from Repeated Interactions},
  shorttitle = {Quantum Collision Models},
  author = {Ciccarello, Francesco and Lorenzo, Salvatore and Giovannetti, Vittorio and Palma, G. Massimo},
  year = {2022},
  month = apr,
  journal = {Phys. Rep.},
  volume = {954},
  pages = {1--70},
  issn = {03701573},
  doi = {10.1016/j.physrep.2022.01.001},
  urldate = {2025-08-19},
  langid = {english}
}

@article{clerk2010,
  title = {Introduction to Quantum Noise, Measurement, and Amplification},
  author = {Clerk, A. A. and Devoret, M. H. and Girvin, S. M. and Marquardt, Florian and Schoelkopf, R. J.},
  year = {2010},
  month = apr,
  journal = {Rev. Mod. Phys.},
  volume = {82},
  number = {2},
  pages = {1155--1208},
  issn = {0034-6861, 1539-0756},
  doi = {10.1103/RevModPhys.82.1155},
  urldate = {2025-08-19},
  langid = {english}
}

@article{cujia2019,
  title = {Tracking the Precession of Single Nuclear Spins by Weak Measurements},
  author = {Cujia, K. S. and Boss, J. M. and Herb, K. and Zopes, J. and Degen, C. L.},
  year = {2019},
  month = jul,
  journal = {Nature},
  volume = {571},
  number = {7764},
  pages = {230--233},
  issn = {0028-0836, 1476-4687},
  doi = {10.1038/s41586-019-1334-9},
  urldate = {2025-08-19},
  langid = {english}
}

@article{deVega2017,
  title = {Dynamics of Non-{{Markovian}} Open Quantum Systems},
  author = {{de Vega}, In{\'e}s and Alonso, Daniel},
  year = {2017},
  month = jan,
  journal = {Rev. Mod. Phys.},
  volume = {89},
  number = {1},
  pages = {015001},
  publisher = {American Physical Society},
  doi = {10.1103/RevModPhys.89.015001}
}

@article{ding2022,
  title = {Non-{{Hermitian}} Topology and Exceptional-Point Geometries},
  author = {Ding, Kun and Fang, Chen and Ma, Guancong},
  year = {2022},
  month = oct,
  journal = {Nat Rev Phys},
  volume = {4},
  number = {12},
  pages = {745--760},
  issn = {2522-5820},
  doi = {10.1038/s42254-022-00516-5},
  urldate = {2025-08-19},
  langid = {english}
}

@article{donvil2023,
  title = {On the {{Unraveling}} of {{Open Quantum Dynamics}}},
  author = {Donvil, Brecht I. C. and {Muratore-Ginanneschi}, Paolo},
  year = {2023},
  month = sep,
  journal = {Open Syst. Inf. Dyn.},
  volume = {30},
  number = {03},
  pages = {2350015},
  issn = {1230-1612, 1793-7191},
  doi = {10.1142/S1230161223500154},
  urldate = {2025-08-19},
  langid = {english}
}

@article{fischer2001observation,
  title = {Observation of the Quantum {{Zeno}} and Anti-{{Zeno}} Effects in an Unstable System},
  author = {Fischer, Martin C and {Guti{\'e}rrez-Medina}, Braulio and Raizen, Mark G},
  year = {2001},
  journal = {Phys. Rev. Lett.},
  volume = {87},
  number = {4},
  pages = {040402},
  publisher = {APS},
  doi = {10.1103/PhysRevLett.87.040402}
}

@article{Gebhart2023,
  title = {Learning Quantum Systems},
  author = {Gebhart, Valentin and Santagati, Raffaele and Gentile, Antonio Andrea and Gauger, Erik M and Craig, David and Ares, Natalia and Banchi, Leonardo and Marquardt, Florian and Pezze, Luca and Bonato, Cristian},
  year = {2023},
  journal = {Nat. Rev. Phys.},
  volume = {5},
  number = {3},
  pages = {141--156},
  publisher = {Nature Publishing Group UK London},
  doi = {10.1038/s42254-022-00552-1}
}

@article{Gilyen2022,
  title = {Quantum Algorithm for Petz Recovery Channels and Pretty Good Measurements},
  author = {Gily{\'e}n, Andr{\'a}s and Lloyd, Seth and Marvian, Iman and Quek, Yihui and Wilde, Mark M.},
  year = {2022},
  month = jun,
  journal = {Phys. Rev. Lett.},
  volume = {128},
  number = {22},
  pages = {220502},
  publisher = {American Physical Society},
  doi = {10.1103/PhysRevLett.128.220502}
}

@article{gneiting2022,
  title = {Unraveling the Topology of Dissipative Quantum Systems},
  author = {Gneiting, Clemens and Koottandavida, Akshay and Rozhkov, A. V. and Nori, Franco},
  year = {2022},
  month = apr,
  journal = {Phys. Rev. Res.},
  volume = {4},
  number = {2},
  pages = {023036},
  issn = {2643-1564},
  doi = {10.1103/PhysRevResearch.4.023036},
  urldate = {2025-08-19},
  langid = {english}
}

@article{gorini1976,
  title = {Completely Positive Dynamical Semigroups of {{{\emph{N}}}} -Level Systems},
  author = {Gorini, Vittorio and Kossakowski, Andrzej and Sudarshan, E. C. G.},
  year = {1976},
  month = may,
  journal = {J. Math. Phys.},
  volume = {17},
  number = {5},
  pages = {821--825},
  issn = {0022-2488, 1089-7658},
  doi = {10.1063/1.522979},
  urldate = {2025-08-19},
  langid = {english}
}

@article{Hamazaki2019,
  title = {Non-{{Hermitian}} Many-Body Localization},
  author = {Hamazaki, Ryusuke and Kawabata, Kohei and Ueda, Masahito},
  year = {2019},
  journal = {Phys. Rev. Lett.},
  volume = {123},
  number = {9},
  pages = {090603},
  publisher = {APS},
  doi = {10.1103/PhysRevLett.123.090603}
}

@article{Hamazaki2020,
  title = {Universality Classes of Non-{{Hermitian}} Random Matrices},
  author = {Hamazaki, Ryusuke and Kawabata, Kohei and Kura, Naoto and Ueda, Masahito},
  year = {2020},
  month = jun,
  journal = {Phys. Rev. Res.},
  volume = {2},
  number = {2},
  pages = {023286},
  publisher = {American Physical Society},
  doi = {10.1103/PhysRevResearch.2.023286}
}

@article{huber2016,
  title = {Topological Mechanics},
  author = {Huber, Sebastian D.},
  year = {2016},
  month = jul,
  journal = {Nat. Phys},
  volume = {12},
  number = {7},
  pages = {621--623},
  issn = {1745-2473, 1745-2481},
  doi = {10.1038/nphys3801},
  urldate = {2025-08-19},
  langid = {english}
}

@article{Jakob2003,
  title = {Variational Functions in Driven Open Quantum Systems},
  author = {Jakob, Matthias and Stenholm, Stig},
  year = {2003},
  month = mar,
  journal = {Phys. Rev. A},
  volume = {67},
  number = {3},
  pages = {032111},
  publisher = {American Physical Society},
  doi = {10.1103/PhysRevA.67.032111}
}

@article{Jin2024,
  title = {Theory of Metastability in Discrete-Time Open Quantum Dynamics},
  author = {Jin, Yuan-De and Qiu, Chu-Dan and Ma, Wen-Long},
  year = {2024},
  month = apr,
  journal = {Phys. Rev. A},
  volume = {109},
  number = {4},
  pages = {042204},
  publisher = {American Physical Society},
  doi = {10.1103/PhysRevA.109.042204}
}

@article{Kawabata2022,
  title = {Many-Body Topology of Non-{{Hermitian}} Systems},
  author = {Kawabata, Kohei and Shiozaki, Ken and Ryu, Shinsei},
  year = {2022},
  month = apr,
  journal = {Phys. Rev. B},
  volume = {105},
  number = {16},
  pages = {165137},
  publisher = {American Physical Society},
  doi = {10.1103/PhysRevB.105.165137}
}

@article{li2013quantum,
  title = {Quantum {{Zeno}} Effect of General Quantum Operations},
  author = {Li, Ying and {Herrera-Mart{\'{\i}}}, David A and Kwek, Leong Chuan},
  year = {2013},
  journal = {Phys. Rev. A},
  volume = {88},
  number = {4},
  pages = {042321},
  publisher = {APS},
  doi = {10.1103/PhysRevA.88.042321}
}

@article{li2019,
  title = {Observation of Parity-Time Symmetry Breaking Transitions in a Dissipative {{Floquet}} System of Ultracold Atoms},
  author = {Li, Jiaming and Harter, Andrew K. and Liu, Ji and De Melo, Leonardo and Joglekar, Yogesh N. and Luo, Le},
  year = {2019},
  month = feb,
  journal = {Nat. Commun.},
  volume = {10},
  number = {1},
  pages = {855},
  issn = {2041-1723},
  doi = {10.1038/s41467-019-08596-1},
  urldate = {2025-08-19},
  langid = {english}
}

@article{lindblad1976,
  title = {On the Generators of Quantum Dynamical Semigroups},
  author = {Lindblad, G.},
  year = {1976},
  month = jun,
  journal = {Commun. Math. Phys.},
  volume = {48},
  number = {2},
  pages = {119--130},
  issn = {0010-3616, 1432-0916},
  doi = {10.1007/BF01608499},
  urldate = {2025-08-19},
  langid = {english}
}

@article{longhi2020,
  title = {Unraveling the Non-{{Hermitian}} Skin Effect in Dissipative Systems},
  author = {Longhi, Stefano},
  year = {2020},
  month = nov,
  journal = {Phys. Rev. B},
  volume = {102},
  number = {20},
  pages = {201103},
  issn = {2469-9950, 2469-9969},
  doi = {10.1103/PhysRevB.102.201103},
  urldate = {2025-08-19},
  langid = {english}
}

@article{Ma2023g,
  title = {Sequential Generalized Measurements: {{Asymptotics}}, Typicality, and Emergent Projective Measurements},
  shorttitle = {Sequential Generalized Measurements},
  author = {Ma, Wen-Long and Li, Shu-Shen and Liu, Ren-Bao},
  year = {2023},
  month = jan,
  journal = {Phys. Rev. A},
  volume = {107},
  number = {1},
  pages = {012217},
  issn = {2469-9926, 2469-9934},
  doi = {10.1103/PhysRevA.107.012217},
  urldate = {2025-08-19},
  langid = {english}
}

@article{Marche2025,
  title = {Exceptional Stationary State in a Dephasing Many-Body Open Quantum System},
  author = {March{\'e}, Alice and Morettini, Gianluca and Mazza, Leonardo and Gotta, Lorenzo and Capizzi, Luca},
  year = {2025},
  month = jul,
  journal = {Phys. Rev. Lett.},
  volume = {135},
  number = {2},
  pages = {020406},
  publisher = {American Physical Society},
  doi = {10.1103/zn9v-k73w}
}

@article{mostafazadeh2002,
  title = {Pseudo-{{Hermiticity}} versus {{PT}} Symmetry: {{The}} Necessary Condition for the Reality of the Spectrum of a Non-{{Hermitian Hamiltonian}}},
  shorttitle = {Pseudo-{{Hermiticity}} versus {{PT}} Symmetry},
  author = {Mostafazadeh, Ali},
  year = {2002},
  month = jan,
  journal = {J. Math. Phys.},
  volume = {43},
  number = {1},
  pages = {205--214},
  issn = {0022-2488, 1089-7658},
  doi = {10.1063/1.1418246},
  urldate = {2025-08-19},
  langid = {english}
}

@article{mostafazadeh2010,
  title = {Pseudo-{{Hermitian Representation}} of {{Quantum Mechanics}}},
  author = {Mostafazadeh, Ali},
  year = {2010},
  month = nov,
  journal = {Int. J. Geom. Methods Mod. Phys.},
  volume = {07},
  number = {07},
  pages = {1191--1306},
  issn = {0219-8878, 1793-6977},
  doi = {10.1142/S0219887810004816},
  urldate = {2025-08-19},
  langid = {english}
}

@article{Nakagawa2021,
  title = {Exact Liouvillian Spectrum of a One-Dimensional Dissipative Hubbard Model},
  author = {Nakagawa, Masaya and Kawakami, Norio and Ueda, Masahito},
  year = {2021},
  month = mar,
  journal = {Phys. Rev. Lett.},
  volume = {126},
  number = {11},
  pages = {110404},
  publisher = {American Physical Society},
  doi = {10.1103/PhysRevLett.126.110404}
}

@article{Nakagawa2025,
  title = {Topology of Discrete Quantum Feedback Control},
  author = {Nakagawa, Masaya and Ueda, Masahito},
  year = {2025},
  month = apr,
  journal = {Phys. Rev. X},
  volume = {15},
  number = {2},
  pages = {021016},
  publisher = {American Physical Society},
  doi = {10.1103/PhysRevX.15.021016}
}

@article{okuma2023,
  title = {Non-{{Hermitian Topological Phenomena}}: {{A Review}}},
  shorttitle = {Non-{{Hermitian Topological Phenomena}}},
  author = {Okuma, Nobuyuki and Sato, Masatoshi},
  year = {2023},
  month = mar,
  journal = {Annu. Rev. Condens. Matter Phys.},
  volume = {14},
  number = {1},
  pages = {83--107},
  issn = {1947-5454, 1947-5462},
  doi = {10.1146/annurev-conmatphys-040521-033133},
  urldate = {2025-08-19},
  langid = {english}
}

@article{ozdemir2019,
  title = {Parity--Time Symmetry and Exceptional Points in Photonics},
  author = {{\"O}zdemir, {\c S}. K. and Rotter, S. and Nori, F. and Yang, L.},
  year = {2019},
  month = aug,
  journal = {Nat. Mater.},
  volume = {18},
  number = {8},
  pages = {783--798},
  issn = {1476-1122, 1476-4660},
  doi = {10.1038/s41563-019-0304-9},
  urldate = {2025-08-19},
  langid = {english}
}

@article{pfender2019,
  title = {High-Resolution Spectroscopy of Single Nuclear Spins via Sequential Weak Measurements},
  author = {Pfender, Matthias and Wang, Ping and Sumiya, Hitoshi and Onoda, Shinobu and Yang, Wen and Dasari, Durga Bhaktavatsala Rao and Neumann, Philipp and Pan, Xin-Yu and Isoya, Junichi and Liu, Ren-Bao and Wrachtrup, J{\"o}rg},
  year = {2019},
  month = feb,
  journal = {Nat. Commun.},
  volume = {10},
  number = {1},
  pages = {594},
  issn = {2041-1723},
  doi = {10.1038/s41467-019-08544-z},
  urldate = {2025-08-19},
  langid = {english}
}

@article{Popkov2021,
  title = {Full Spectrum of the Liouvillian of Open Dissipative Quantum Systems in the Zeno Limit},
  author = {Popkov, Vladislav and Presilla, Carlo},
  year = {2021},
  month = may,
  journal = {Phys. Rev. Lett.},
  volume = {126},
  number = {19},
  pages = {190402},
  publisher = {American Physical Society},
  doi = {10.1103/PhysRevLett.126.190402}
}

@article{Qiu2024,
  title = {How Coherence Measurements of a Qubit Steer Its Quantum Environment},
  author = {Qiu, Chu-Dan and Jin, Yuan-De and Zhang, Jun-Xiang and Liu, Gang-Qin and Ma, Wen-Long},
  year = {2024},
  month = jul,
  journal = {Phys. Rev. B},
  volume = {110},
  number = {2},
  pages = {024311},
  publisher = {American Physical Society},
  doi = {10.1103/PhysRevB.110.024311}
}

@article{roccati2022,
  title = {Non-{{Hermitian Physics}} and {{Master Equations}}},
  author = {Roccati, Federico and Palma, G. Massimo and Ciccarello, Francesco and Bagarello, Fabio},
  year = {2022},
  month = mar,
  journal = {Open Syst. Inf. Dyn.},
  volume = {29},
  number = {01},
  pages = {2250004},
  issn = {1230-1612, 1793-7191},
  doi = {10.1142/S1230161222500044},
  urldate = {2025-08-19},
  langid = {english}
}

@article{rotter2009,
  title = {A Non-{{Hermitian Hamilton}} Operator and the Physics of Open Quantum Systems},
  author = {Rotter, Ingrid},
  year = {2009},
  month = apr,
  journal = {J. Phys. A},
  volume = {42},
  number = {15},
  pages = {153001},
  issn = {1751-8113, 1751-8121},
  doi = {10.1088/1751-8113/42/15/153001},
  urldate = {2025-08-19}
}

@article{Sarkar1995,
  title = {Using the Matrix Pencil Method to Estimate the Parameters of a Sum of Complex Exponentials},
  author = {Sarkar, T.K. and Pereira, O.},
  year = {1995},
  month = feb,
  journal = {IEEE Antennas Propag. Mag.},
  volume = {37},
  number = {1},
  pages = {48--55},
  issn = {1045-9243},
  doi = {10.1109/74.370583},
  urldate = {2025-08-19},
  langid = {english}
}

@article{Scarani2002,
  title = {Thermalizing Quantum Machines: {{Dissipation}} and Entanglement},
  author = {Scarani, Valerio and Ziman, M{\'a}rio and {{\v S} {\v S}telmachovi{\v c} {\v c}}, Peter and Gisin, Nicolas and {Bu{\v z} {\v z}ek}, Vladim{\'{\i}}r},
  year = {2002},
  month = feb,
  journal = {Phys. Rev. Lett.},
  volume = {88},
  number = {9},
  pages = {097905},
  publisher = {American Physical Society},
  doi = {10.1103/PhysRevLett.88.097905}
}

@article{Stenholm2002,
  title = {Variational Functions in Open Systems},
  author = {Stenholm, Stig},
  year = {2002},
  journal = {Ann. Phys.},
  volume = {302},
  number = {2},
  pages = {142--157},
  publisher = {Elsevier},
  doi = {10.1006/aphy.2002.6309}
}

@article{Stenholm2004,
  title = {Time Inversion in Dynamical Systems},
  author = {Stenholm, Stig and Jakob, Matthias},
  year = {2004},
  journal = {Ann. Phys.},
  volume = {310},
  number = {1},
  pages = {106--126},
  publisher = {Elsevier},
  doi = {10.1016/j.aop.2003.09.001}
}

@article{takata2018,
  title = {Photonic {{Topological Insulating Phase Induced Solely}} by {{Gain}} and {{Loss}}},
  author = {Takata, Kenta and Notomi, Masaya},
  year = {2018},
  month = nov,
  journal = {Phys. Rev. Lett.},
  volume = {121},
  number = {21},
  pages = {213902},
  issn = {0031-9007, 1079-7114},
  doi = {10.1103/PhysRevLett.121.213902},
  urldate = {2025-08-19},
  langid = {english}
}

@article{wang2021,
  title = {Topological Physics of Non-{{Hermitian}} Optics and Photonics: A Review},
  shorttitle = {Topological Physics of Non-{{Hermitian}} Optics and Photonics},
  author = {Wang, Hongfei and Zhang, Xiujuan and Hua, Jinguo and Lei, Dangyuan and Lu, Minghui and Chen, Yanfeng},
  year = {2021},
  month = dec,
  journal = {J. Opt.},
  volume = {23},
  number = {12},
  pages = {123001},
  issn = {2040-8978, 2040-8986},
  doi = {10.1088/2040-8986/ac2e15},
  urldate = {2025-08-19}
}

@article{wangNonHermitianOpticsPhotonics2023,
  title = {Non-{{Hermitian}} Optics and Photonics: {{From}} Classical to Quantum},
  shorttitle = {Non-{{Hermitian}} Optics and Photonics},
  author = {Wang, Changqing and Fu, Zhoutian and Mao, Wenbo and Qie, Jinran and Stone, A. Douglas and Yang, Lan},
  year = {2023},
  month = jun,
  journal = {Adv. Opt. Photon.},
  volume = {15},
  number = {2},
  pages = {442},
  issn = {1943-8206},
  doi = {10.1364/AOP.475477},
  urldate = {2025-08-19},
  langid = {english}
}

@book{watrous2018,
  title = {The {{Theory}} of {{Quantum Information}}},
  author = {Watrous, John},
  year = {2018},
  month = apr,
  edition = {1},
  publisher = {Cambridge University Press},
  doi = {10.1017/9781316848142},
  urldate = {2025-08-19},
  isbn = {978-1-316-84814-2 978-1-107-18056-7}
}

@book{wiseman2009,
  title = {Quantum {{Measurement}} and {{Control}}},
  author = {Wiseman, Howard M. and Milburn, Gerard J.},
  year = {2009},
  month = nov,
  edition = {1},
  publisher = {Cambridge University Press},
  doi = {10.1017/CBO9780511813948},
  urldate = {2025-08-19},
  isbn = {978-0-521-80442-4 978-0-511-81394-8 978-1-107-42415-9}
}

@book{wolf2012,
  title = {Quantum Channels and Operations-Guided Tour},
  author = {Wolf, Michael M},
  year = {2012}
}

@article{Yao2018,
  title = {Edge States and Topological Invariants of Non-Hermitian Systems},
  author = {Yao, Shunyu and Wang, Zhong},
  year = {2018},
  month = aug,
  journal = {Phys. Rev. Lett.},
  volume = {121},
  number = {8},
  pages = {086803},
  publisher = {American Physical Society},
  doi = {10.1103/PhysRevLett.121.086803}
}

@article{Yokomizo2019,
  title = {Non-Bloch Band Theory of Non-Hermitian Systems},
  author = {Yokomizo, Kazuki and Murakami, Shuichi},
  year = {2019},
  month = aug,
  journal = {Phys. Rev. Lett.},
  volume = {123},
  number = {6},
  pages = {066404},
  publisher = {American Physical Society},
  doi = {10.1103/PhysRevLett.123.066404}
}

@article{yoshida2019,
  title = {Exceptional Rings Protected by Emergent Symmetry for Mechanical Systems},
  author = {Yoshida, Tsuneya and Hatsugai, Yasuhiro},
  year = {2019},
  month = aug,
  journal = {Phys. Rev. B},
  volume = {100},
  number = {5},
  pages = {054109},
  issn = {2469-9950, 2469-9969},
  doi = {10.1103/PhysRevB.100.054109},
  urldate = {2025-08-19},
  langid = {english}
}

@article{Zhang2024,
  title={Observation of metastability in open quantum dynamics of a solid-state system},
  author={Zhang, Jun-Xiang and Jin, Yuan-De and Qiu, Chu-Dan and Ma, Wen-Long and Liu, Gang-Qin},
  journal={Nat. Commun.},
  volume={16},
  number={1},
  pages={9818},
  year={2025},
  publisher={Nature Publishing Group UK London},
  url={https://www.nature.com/articles/s41467-025-64772-6}
}

@article{Zheng2025,
  title = {Near-Optimal Performance of Quantum Error Correction Codes},
  author = {Zheng, Guo and He, Wenhao and Lee, Gideon and Jiang, Liang},
  year = {2024},
  month = jun,
  journal = {Phys. Rev. Lett.},
  volume = {132},
  number = {25},
  pages = {250602},
  publisher = {American Physical Society},
  doi = {10.1103/PhysRevLett.132.250602}
}

@article{Lange2011,
  title = {Single-Spin Magnetometry with Multipulse Sensing Sequences},
  author = {de Lange, G. and Rist\`e, D. and Dobrovitski, V. V. and Hanson, R.},
  journal = {Phys. Rev. Lett.},
  volume = {106},
  issue = {8},
  pages = {080802},
  numpages = {4},
  year = {2011},
  month = {Feb},
  publisher = {American Physical Society},
  doi = {10.1103/PhysRevLett.106.080802},
  url = {https://link.aps.org/doi/10.1103/PhysRevLett.106.080802}
}

@article{Mamin2013,
author = {H. J. Mamin  and M. Kim  and M. H. Sherwood  and C. T. Rettner  and K. Ohno  and D. D. Awschalom  and D. Rugar },
title = {Nanoscale Nuclear Magnetic Resonance with a Nitrogen-Vacancy Spin Sensor},
journal = {Science},
volume = {339},
number = {6119},
pages = {557-560},
year = {2013},
doi = {10.1126/science.1231540}}

@article{Loretz2013,
  title = {Radio-Frequency Magnetometry Using a Single Electron Spin},
  author = {Loretz, M. and Rosskopf, T. and Degen, C. L.},
  journal = {Phys. Rev. Lett.},
  volume = {110},
  issue = {1},
  pages = {017602},
  numpages = {5},
  year = {2013},
  month = {Jan},
  publisher = {American Physical Society},
  doi = {10.1103/PhysRevLett.110.017602},
  url = {https://link.aps.org/doi/10.1103/PhysRevLett.110.017602}
}

@article{Abobeih2019,
  title={Atomic-scale imaging of a 27-nuclear-spin cluster using a quantum sensor},
  author={Abobeih, MH and Randall, J and Bradley, CE and Bartling, HP and Bakker, MA and Degen, MJ and Markham, M and Twitchen, DJ and Taminiau, TH},
  journal={Nature},
  volume={576},
  number={7787},
  pages={411--415},
  year={2019},
  publisher={Nature Publishing Group UK London},
  doi={10.1038/s41586-019-1834-7}
}

@article{Bradley2019,
  title = {A Ten-Qubit Solid-State Spin Register with Quantum Memory up to One Minute},
  author = {Bradley, C. E. and Randall, J. and Abobeih, M. H. and Berrevoets, R. C. and Degen, M. J. and Bakker, M. A. and Markham, M. and Twitchen, D. J. and Taminiau, T. H.},
  journal = {Phys. Rev. X},
  volume = {9},
  issue = {3},
  pages = {031045},
  numpages = {12},
  year = {2019},
  month = {Sep},
  publisher = {American Physical Society},
  doi = {10.1103/PhysRevX.9.031045},
  url = {https://link.aps.org/doi/10.1103/PhysRevX.9.031045}
}

@article{Taminiau2012,
  title = {Detection and Control of Individual Nuclear Spins Using a Weakly Coupled Electron Spin},
  author = {Taminiau, T. H. and Wagenaar, J. J. T. and van der Sar, T. and Jelezko, F. and Dobrovitski, V. V. and Hanson, R.},
  journal = {Phys. Rev. Lett.},
  volume = {109},
  issue = {13},
  pages = {137602},
  numpages = {5},
  year = {2012},
  month = {Sep},
  publisher = {American Physical Society},
  doi = {10.1103/PhysRevLett.109.137602},
  url = {https://link.aps.org/doi/10.1103/PhysRevLett.109.137602}
}

@article{hoeffding1963probability,
  title={Probability inequalities for sums of bounded random variables},
  author={Hoeffding, Wassily},
  journal={J. Am. Stat. Assoc.},
  volume={58},
  number={301},
  pages={13--30},
  year={1963},
  doi={10.2307/2282952},
  publisher={Taylor \& Francis}
}

\onecolumngrid % Switch to one column layout for SM (optional)

\clearpage

%\begin{center}
%	\textbf{Supplementary Note I - Proof on the pseudo-hermiticity of quantum channel.}	
%\end{center}
%Given a quantum channel $\phii$ with spectral decomposition
%\begin{equation}
%  \phii=\sum_j \lambda_j \kett{R_j}\braa{L_j},
%\end{equation}
%we construct a metric operator 
%\begin{equation}
%  \eta=\sum_{\{\lambda_j\}\in\mathbb{R}}\kett{L_j}\braa{L_j}+\sum_{\%{\lambda_k\}\in\mathbb{C}}\kett{L_k}\braa{L_k'},
%\end{equation}
%in which $\kett{L_k'}$ denotes the left eigenstate corresponding to the eigenvalue $\lambda_k^*$ since the eigenvalues appear in complex pairs. 

%Then 
%\begin{equation}
%  \eta^{-1}=\sum_{\{\lambda_j\}\in\mathbb{R}}\kett{R_j}\braa{R_j}+\sum_{\{\lambda_k\}\in\mathbb{C}}\kett{R_k'}\braa{R_k},
%\end{equation}
%and
%\begin{equation}
%\begin{aligned}
%	    \eta\phii\eta^{-1}&=\sum_{\{\lambda_j\}\in\mathbb{R}}\lambda_j \kett{L_j}\braa{R_j}+\sum_{\{\lambda_k\}\in\mathbb{C}}\lambda_k \kett{L_k'}\braa{R_k'}	\\
%	    &=\sum_{\{\lambda_j\}\in\mathbb{R}}\lambda_j \kett{L_j}\braa{R_j}+\sum_{\{\lambda_k\}\in\mathbb{C}}\lambda_k^* \kett{L_k}\braa{R_k}	\\
%	    &=\sum_{j}\lambda_j^* \kett{L_j}\braa{R_j}\\
%	    &=\phii^\dagger.
%\end{aligned}
%\end{equation}
%Then, the constructed operator $\eta$ is a metric operator, and $\phii$ is pseudo-Hermitian. One can also easily verify $\eta$ is an Hermitian operator, i.e. $\eta=\eta^\dagger$.

%\begin{center}
%    SI is separated into a single file.
%\end{center}

\end{document}

%\begin{CJK*}{UTF8}{gbsn}
%\linenumbers

%\title{Non-Hermitian Physics in Quantum Channels: Pseudo-Hermiticity, Spectrum Estimation and Exceptional Points}
\title{Supplementary Material for ``Spectrum measurement of quantum channels and application to Hamiltonian parameter estimation''}
\author{Yuan-De Jin}
\affiliation{State Key Laboratory of Semiconductor Physics and Chip Technologies, Institute of Semiconductors, Chinese Academy of Sciences, Beijing 100083, China}
\affiliation{Center of Materials Science and Opto-Electronic Technology, University of Chinese Academy of Sciences, Beijing 100049, China}

\author{Wen-Long Ma}
\email{wenlongma@semi.ac.cn}
\affiliation{State Key Laboratory of Semiconductor Physics and Chip Technologies, Institute of Semiconductors, Chinese Academy of Sciences, Beijing 100083, China}
\affiliation{Center of Materials Science and Opto-Electronic Technology, University of Chinese Academy of Sciences, Beijing 100049, China}
\date{\today }
\begin{abstract}

%generated by a free Hamiltonian on a target quantum system

\end{abstract}
\maketitle
\tableofcontents

%\section{Details in non-Hermitian properties of quantum channel}
\section{Proof on the pseudo-Hermiticity of Hermitian-preserving maps}
\textbf{Theorem 1}.
Let $\mcm(\cdot)$ be a Hermitian-preserving linear map, i.e., $\mcm(X)^\dagger=\mcm(X)$ for any Hermitian operator $X$. If it is diagonalizable with a discrete spectrum in the vectorized operator space as
\begin{equation}
 \hat\mcm=\sum_j \lambda_j \kett{R_j}\braa{L_j},
\end{equation}
where $\{\kett{R_j},\kett{L_j}\}$ is a complete biorthonormal basis satisfying $\brakett{L_i}{R_j}=\delta_{ij}$ with $\delta_{ij}$ being the Kronecker delta. Then such a map is a pseudo-Hermitian operator on the vectorized operator space.
\begin{proof}
According to Ref.~\cite{mostafazadeh2002}, the operator $\hat\mcm$ on the vectorized operator space is $\eta$-pseudo-Hermitian, i.e., there exists a Hermitian and invertible metric operator $\eta$ making $\eta\mm\eta^{-1}=\mm^\dagger$, if and only if one of the following conditions holds
\begin{enumerate}

    \item The spectrum of $\hat\mcm$ is real, then $\mm$ is $\mathbb{I}$-pseudo-Hermitian or Hermitian.
    \item The complex eigenvalues come in complex conjugate pairs and the multiplicities of complex conjugate eigenvalues are the same.
\end{enumerate}

A linear map is Hermitian-preserving if and only if $\mcm(X)^\dagger=\mcm(X^\dagger)$ \cite{watrous2018}, since
\begin{equation}
    \mcm(X^\dagger)=\mcm(H)-i\mcm(A)=\mcm(X)^\dagger,
\end{equation}
where we decompose $X$ into Hermitian and anti-Hermitian parts, i.e., $X=H+iA$. Then for any complex eigenvalue $\lambda_j$ with right eigenmatrix $R_j$, we have $\mcm(R_j)^\dagger=\mcm(R_j^\dagger)=\lambda_j^*R_j^\dagger$, so $R_j^\dagger$ is the right eigenmatrix for eigenvalue $\lambda_j^*$. This means that the eigenvalues always appear in complex-conjugate pairs, and the multiplicities of complex-conjugate eigenvalues are the same. Thus, any Hermitian-preserving linear map is a pseudo-Hermitian operator on the vectorized operator space if it is diagonalizable.
\end{proof}

There are two typical cases of the Hermitian-preserving map, i.e., the Liouvillians (or the Lindblad superoperators), and the complete positive (CP) maps.

The Liouvillian is defined by the Lindblad master equation as
\begin{equation}
    \mathcal{L}(\cdot)=-i[H,(\cdot)]+\sum_l\left[L_l(\cdot) L_l^\dagger-\frac{1}{2}\left(L_l^\dagger L_l(\cdot)+(\cdot) L_l^\dagger L_l\right)\right],
\end{equation}
from which we can directly verify its Hermitian-preserving property, since $\mathcal{L}(X)^{\dagger}=\mathcal{L}(X^{\dagger})$.
%\begin{equation}
%    \mcl(\rho)^\dagger=i[\rho,H]+\sum_l\left[ L_l\rho L_l^\dagger-\frac{1}{2}\left(\rho L_l^\dagger L_l+L_l^\dagger %L_l\rho\right)\right]=\mcl(\rho).
%\end{equation}

According to the Kraus theorem, the CP map $\mcm$ has the general form
\begin{equation}
    \mcm(\cdot)=\sum_{i=1}^r M_i(\cdot)M_i^\dagger,
\end{equation}
which is also Hermitian-preserving since $\mcm(\cdot)^\dagger=\sum_{i=1}^r M_i(\cdot)M_i^\dagger=\mcm(\cdot)$. We note that a quantum channel, as a special CP map satisfying  the trace-preserving condition ($\sum_{i=1}^r M_i^\dagger M_i=\mathbb{I}$), is also pseudo-Hermitian if it is diagonalizable in the vectorized operator space.

For the Hermitian-preserving map $\mcm$, we can explicitly construct the metric operator $\eta$ as \cite{Stenholm2002,Stenholm2004,Jakob2003} 
\begin{equation}
 \eta=\sum_{\{\lambda_j\}\in\mathbb{R}}a_j\kett{L_j}\braa{L_j}+\sum_{\{\lambda_k\}\in\mathbb{C/R}}\kett{L_k}\langle\langle{L_k^\dagger}|,
\end{equation}
with $a_j\in\{-1,1\}$. Then its inverse is
\begin{equation}
 \eta^{-1}=\sum_{\{\lambda_j\}\in\mathbb{R}}a_j\kett{R_j}\braa{R_j}+\sum_{\{\lambda_k\}\in\mathbb{C/R}}\kett{R_k^\dagger}\braa{R_k}.
\end{equation}
One can easily verify that $\eta\hat\mcm\eta^{-1}=\hat\mcm^\dagger$.

We can also characterize the channel by another symmetry, which we call the swap-time symmetry. We define the swap-time symmetry operator as $\hat\mct=\hat S\hat K$, where $\hat K$ is the complex conjugation and $\hat S$ is the swap operation on the vectorized operator space, i.e., $\hat S\kett{ij}=\kett{ji}$. Note that $\hat K^{-1}=\hat K$ and $\hat S^{-1}=\hat S$. Then the channel has the swap-time symmetry since $\hat\mct \phii\hat\mct^{-1}=\sum_{i}\hat S(M_i^*\otimes M_i)\hat S^{-1}=\phii$. Moreover, any eigenvector $\kett{R_j}$ of the channel with a non-degenerate real eigenvalue also have such a symmetry, since $\hat\mct \kett{R_j}=\kett{R_j^{\dagger}}=\kett{R_j}$. However, the eigenvectors with complex conjugate eigenvalues often breaks such a symmetry since typically $\hat\mct\kett{R_k}=\kett{R_k^\dagger}\neq \ee^{\ii  \theta}|R_k\rangle\rangle$ with $\ee^{\ii  \theta}$ being an arbitrary phase.

\section{Details in channel spectrum measurement}
\subsection{Measurement statistics for non-diagonalizable quantum channels}\label{Sec:nonDiagonal}
%\subsection{Jordan decomposition of the quantum channel}
In the main text, we study the measurement statistics for diagonalizable quantum channels. We can further extend those results to any non-diagonalizable channel, which can be decomposed into a direct sum of Jordan normal form by a similarity transformation
\begin{equation}
    \phii=\mathcal{S}\left(\bigoplus_{j=1}^K \mathcal{J}_{d_k}(\lambda_k)\right)\mathcal{S}^{-1},
\end{equation}
where $\mathcal S$ is the similarity transformation matrix, $\mathcal{J}_{d_k}(\lambda_k)=\lambda_k \mathbb{I}_{d_k}+N_{k}$ is a $d_k$-dimensional Jordan block with the eigenvalue $\lambda_k$, and  $N_k$ is the nilpotent part of the block, represented as an upper-triangular matrix with ones on the superdiagonal and zeros elsewhere, satisfying $N_k^{\,d_k}=0$. Here $\sum_{k=1}^Kd_k=d^2$, where $d$ is the dimension of the system. When $d_k=1$ for $k=1,...,K$ (and thus $K=d$), the channel can be diagonalized. We note that the Jordan blocks for fixed points are trivial, i.e. $d_k=1$ for $\lambda_k=1$, then the channel can be further decomposed as 
\begin{equation}
        \phii=\mathcal{S}\left[\sum_{\lambda_k=1}\mathcal{P}_k+\sum_{\lambda_k\neq 1 }(\lambda_k\mcp_k+\mcn_k)\right]\mathcal{S}^{-1},
\end{equation}
where $\mcp_k$ denotes the projection onto the Jordan block sufficing $\mcp_k\mcp_l=\delta_{kl}\mcp_l$ with $\lambda_k$, $\mcn_k$ is the nilpotent part on that subspace, satisfying $\mcn_k^{d_k}=0$ and $\mcp_k \mcn_k = \mcn_k \mcp_k = \mcn_k$.

Then $m$ repetitive channels can be expressed as% (below we ignore the similarity transformation matrix)
\begin{equation}
\begin{aligned}
        \phii^m=&\mcs \left[\sum_{\lambda_k=1}\mcp_k+\sum_{\lambda_k\neq1}(\lambda_k\mcp_k+\mcn_k)^m \right ]\mcs^{-1} \\
          =&\mcs\left[\sum_{\lambda_k=1}\mcp_k+\sum_{\lambda_k\neq1}\sum_{r=0}^{d_k-1} \binom{m}{r}\lambda_k^{m-r}\mcn_k^{r}\mcp_k \right ]\mcs^{-1}.
\end{aligned}
\end{equation}
Then we consider the probability for obtaining outcome $i$ in the $(m+1)$-th measurement cycle is
\begin{equation}
\begin{aligned}
    p_i^{m+1}&=\braa{\mathbb{I}}\mm_i \phii^{m}\kett{\rho}\\
    &=\braa{\mathbb{I}}\mm_i\mcs\left[\sum_{\lambda_k=1}\mcp_k+\sum_{\lambda_k\neq1}\sum_{r=0}^{d_k-1} \binom{m}{r}\lambda_k^{m-r}\mcn_k^{r}\mcp_k\right]\mcs^{-1}\kett{\rho}\\
    &=\sum_{\lambda_k=1}\braa{\mathbb{I}}\mm_i\mcs\mcp_k \mcs^{-1}\kett{\rho}+\sum_{\lambda_k\neq1}\lambda_k^m\sum_{r=0}^{d_k-1} \binom{m}{r}\lambda_k^{-r}\braa{\mathbb{I}}\mm_i\mcs\mcn_k^{r}\mcp_k\mcs^{-1}\kett{\rho}\\
    &=\sum_{\lambda_k=1}c_{k,0}+\sum_{\lambda_k\neq1}\lambda_k^m\sum_{r=0}^{d_k-1} \binom{m}{r}\lambda_k^{-r} c_{k,r},
\end{aligned}
\label{Eq:probJordan}
\end{equation}
with $c_{k,r}=\braa{\mathbb{I}}\mm_i\mcs\mcn_k^{r}\mcp_k \mcs^{-1}\kett{\rho}$.

For a channel with some second-order exceptional points (EPs), Eq.~\eqref{Eq:probJordan} becomes 
\begin{equation}
\begin{aligned}
p_i^{m+1}=\sum_{k}c_{k,0}\lambda_k^m+\sum_{d_k=2}c_{k,1} m\lambda_k^{m-1} ,
\end{aligned}
\label{Eq:probSecOrdEP}
\end{equation}
which contains some exponential polynomial terms besides the original pure exponential term. 

\begin{figure}
    \centering
    \includegraphics[width=\linewidth]{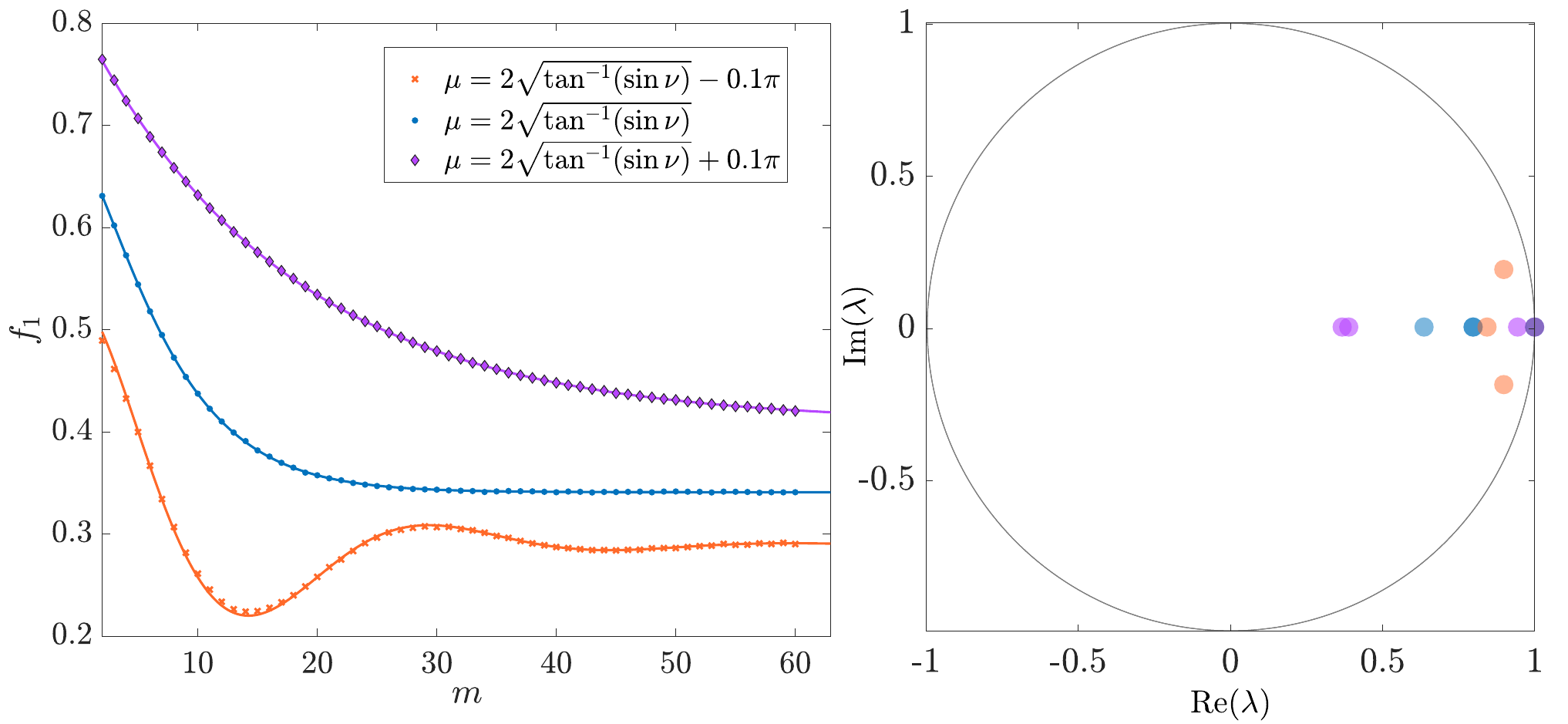}
    \caption{The frequency $f_1$ as a function of measurement cycle $m$ below (orange crosses), at (blue points) and beyond (purple triangles) the EP. The data below, at and beyond the EP are fitted by $f_1^{(m)}=c_{1,0}+2\Re (c_{2,0})|\lambda_2|^m\cos(m\varphi_2)$ (orange line), $f_1^{(m)}=c_{1,0}+c_{2,0}\lambda_2^m+c_{2,1}m\lambda_2^{m-1}$ (blue line) and $f_1^{(m)}=c_{1,0}+c_{2,0}\lambda_2^m$ (purple line), respectively.  The corresponding eigenvalues are shown in the right, near the EP, a pair of conjugate eigenvalues (orange points) coalesce (blue points) and transform into two real eigenvalues (purple points).}
    %The simulation data (points) are fitted by $c_{1,0}+c_{2,0}\lambda_2^m+c_{2,1}m\lambda_2^{m-1}$ (blue line, for the data at EP),  $c_{1,0}+c_{2,0}\lambda_2^m+c_{3,0}\lambda_3^m$ (purple line, for the purely damping data) and $c_{1,0}+2\Re (c_{2,0})|\lambda_2|^m\cos(m\varphi_2)$ (orange line, for the damped oscillation data). 

    \label{fig:f1EP}
\end{figure}

Below we show a numerical simulation under the exactly solvable model in Example I of the main text, in which the coupling Hamiltonian is $A=g\sigma_x /2$, and $B=\omega\sigma_z/2$ (Fig.~\ref{fig:f1EP}). This model contains a second-order EP line at  $\tan^4\left(\frac{\mu}{2}\right)=\sin^2\nu$ in the $(\mu,\nu)$ plane of the parameter space, where we define two parameters $\mu=g\tau_A$ and $\nu=\omega \tau_B$. One can see that the measurement statistics for the parameter $\mu$ at the EP ($\mu=2\sqrt{\tan^{-1}(\sin\nu)}$) can be well described by Eq. \eqref{Eq:probSecOrdEP}. Moreover, the results below the EP ($\mu=2\sqrt{\tan^{-1}(\sin\nu)}-0.1\pi$) and beyond the EP ($\mu=2\sqrt{\tan^{-1}(\sin\nu)}+0.1\pi$) clearly show the transition from damped oscillations to exponential decays.  
%Here we note that at an EP point, $\lambda_2=\lambda_3\in\mathbb{R}$, and in our system, $c_{4,0}=0$.

\subsection{Measurement statistics for correlation measurements}\label{Sec:corrFunc}
In the main text, we measure the channel spectrum by tracking the measurement statistics of a specific outcome in sequential quantum channels. Here we analyze the statistics for the two-point correlation function $C(m)=\langle \alpha_{m+1}\alpha_1\rangle$, where $\alpha_k\in\{0,1\}$ is the measurement outcome of the $k$th cycle. We can derive the correlation function as 
\begin{equation}
\begin{aligned}
C(m) %&=\sum_{\alpha_{m+1},\alpha_1}\alpha_{m+1}\alpha_1\braa{\I}\mm_{\alpha_{m+1}}\phii^{m-1}\mm_{\alpha_1}\kett{\rho}\\
=\braa{\I}\mm_{1}\phii^{m-1}\mm_{1}\kett{\rho} %\\
=\sum_{j} d_j\lambda_j^{m-1},
   \end{aligned} 
\end{equation}
with $d_j= \braa{\I}\mm_{1} \kett{R_j}\braa{L_j}\mm_{1}\kett{\rho}$. So the statistics of correlation measurements can also be expressed by a sum of exponential functions of the channel spectrum as in the direct tracking of measurement outcomes, differing only in the coefficients.

Similarly, if we relabel the measurement outcome of the $k$th cycle as $\alpha_k\in\{+1,-1\}$, the correlation function becomes
\begin{equation}
    \begin{aligned}
C(m) %&=\sum_{\alpha_{m+1},\alpha_1}\alpha_{m+1}\alpha_1\braa{\I}\mm_{\alpha_{m+1}}\phii_B\phii^{m-1}\mm_{\alpha_1}\kett{\rho}\\ 
=\braa{\I}\hat\mcp\phii^{m-1}\hat\mcp\kett{\rho}%\\
=\sum_{j} g_j\lambda_j^{m-1}
\end{aligned}
\end{equation}
where $\hat\mcp=\mm_0-\mm_1$ and $g_j=\braa{\I}\hat\mcp \kett{R_j}\braa{L_j}\hat\mcp\kett{\rho}$.
\subsection{Basics of matrix pencil (MP) method}
Here we briefly summarize the MP method. MP method is an approach to obtain the poles $\{z_j\}$ of signals $y(m)$ modeled as a superposition of complex exponentials
\begin{equation}
    y(m)=\sum_{j=1}^Jc_j z_j^m+n(m),\quad m=1,...,N,
\end{equation}
where $|z_j|<1$, and $n(m)$ is a noise function. For the case without noise, two $(N-L)\times L$ matrices can be defined as
\begin{equation}
    Y_1 =
\begin{pmatrix}
y(1) & y(2) & \cdots & y(L) \\
y(2) & y(3) & \cdots & y(L+1) \\
\vdots & \vdots & \ddots & \vdots \\
y(N-L) & y(N-L+1) & \cdots & y(N-1)
\end{pmatrix}, \qquad
Y_2 =
\begin{pmatrix}
y(2) & y(3) & \cdots & y(L+1) \\
y(3) & y(4) & \cdots & y(L+2) \\
\vdots & \vdots & \ddots & \vdots \\
y(N-L+1) & y(N-L+2) & \cdots & y(N),
\end{pmatrix}
\end{equation}
where $L$ is the pencil length. Then the poles are the generalized eigenvalues of the matrix pair $(Y_2,Y_1)$, i.e.,
\begin{equation}
    Y_2v_j=z_j Y_1 v_j.
\end{equation}
Equivalently, $\{z_j\}$ are also the eigenvalues of the matrix $(Y_1^\dagger Y_1)^{-1}Y_1^\dagger Y_2$. For the case with noise, one can define an $(N-L+1)\times L$ dimensional matrix
\begin{equation}
    Y =
\begin{pmatrix}
y(1) & y(2) & \cdots & y(L+1) \\
y(2) & y(3) & \cdots & y(L+2) \\
\vdots & \vdots & \ddots & \vdots \\
y(N-L) & y(N-L+1) & \cdots & y(N).
\end{pmatrix}
\end{equation}
Then we construct a singular-value decomposition of the matrix $Y$, i.e., $Y= U\Sigma V^\dagger$, and replace $Y$ with $Y'=U\Sigma' V^\dagger$ and $\Sigma'$ only contains the largest $J$ singular values. Then $Y_1'$ and $Y_2'$ can be selected by deleting the  last column and the first column of $Y'$ to calculate the generalized eigenvalues. The pencil length is usually selected to be $N/3\sim N/2$ to suppress the noise.
%Specifically, the correlation function is analyzed in Ref. \cite{pfender2019} for the single qubit target system in the main text and Sec.~\ref{Sec:nonDiagonal} with $A=g\sigma_x/2$ and $B=\omega\sigma_z/2$. By observing the pattern of the correlation function, they obtain the corresponding eigenvalues for the system.

% \begin{equation}
%     C(m)=\langle \alpha_{m+1} \alpha_{1}\rangle=c^{(\pm)}_{12}\lambda_{12}^m+c^{(\pm)}_{21}\lambda_{21}^m,
% \end{equation}
% where $\alpha=\pm 1$ is taken, $c^{(\pm)}_{12,21}=\frac{\sin^2\mu}{2}\left(1\pm\tan^2\frac{\mu}{2}\frac{\cos\nu}{\sqrt{\tan^4\frac{\mu}{2}-\sin^2\nu}}\right)$ and $\lambda_{12,21}=\cos^2\left(\frac{\mu}{2}\right)[\cos\nu\pm\sqrt{\tan^4\left(\frac{\mu}{2}\right)-\sin^2\nu}]$ are the same with those introduced in the main text. We note that $\lambda_{11}=\lambda_{22}=0$ is given here. 
%\section{Details in quantum channel spectrum estimation}

\section{Details in Hamiltonian parameter estimation}
\subsection{Deriving the spectrum of the concatenated channel from perturbation theory}
The unitary channel $\phii_B$ generated by the free Hamiltonian $B=\sum_i b_i\ket{i}\bra{i}$ can be expanded in the vectorized operator space as 
\begin{equation}
    \phii_B=V\otimes V^*=\sum_{ij}v_{ij}\kett{ij}\braa{ij},
\end{equation}
with $v_{ij}=\ee^{-\ii \bij\tau_B}$ and $\bij=b_i-b_j$. The channel $\phii_A$ induced by the RIM can be expressed as $\phii_A=\sum_{\alpha={0,1}}M_\alpha\otimes M_\alpha^*$
where $M_\alpha=[U_0-(-1)^{\alpha} \ee^{\ii\phi}U_1]/2$ and $U_{\alpha}=\exp{-i(-1)^\alpha A \tau_A }$.
We note that
\begin{equation}
\begin{aligned}
        U_0\otimes U_0^*=\ee^{-\ii A\tau_A}\otimes \ee^{\ii  A\tau_A}%\\
        =(\ee^{-\ii A\tau_A\otimes \mathbb{I}})(\ee^{\mathbb{I}\otimes iA\tau_A})%\\
        =\ee^{-\ii \tau_A\hat\mcc_A},
\end{aligned}
\end{equation}
where $\hat{\mathcal C}_A=A\otimes \mathbb{I}-\mathbb{I}\otimes A^T$ is the superoperator of the commutator $[A,\cdot]=A(\cdot) \mathbb{I}-\mathbb{I}(\cdot) A$ giving $\hat{\mathcal C}_A|Y\rangle\rangle=|[A,Y]\rangle\rangle$. Similarly, we have $U_1\otimes U_1^*=\ee^{\ii  \tau_A\hat\mcc_A}$, and thus
\begin{equation}
    \phii_A=\frac{1}{2}(U_0\otimes U_0^*+U_1\otimes U_1^*)=\cos(\tau_A\hat\mcc_A),
\end{equation}
When $||\mcc_A||\tau_A\ll 1$, we can expand $\phii_A$ and retain terms up to the second order $\tau_A^2$,
\begin{equation}
    \phii_A\approx\mathbb{I}\otimes \mathbb{I}-\frac{1}{2}\tau_A^2\mcc_A^2  =\mathbb{I}\otimes \mathbb{I}+\tau_A^2\hat{ \mathcal{L}},
\end{equation}
where $\hat{ \mathcal{L}}=-\hat\mcc_A^2/2=A\otimes A^T-\frac{1}{2}[A^2\otimes \mathbb{I}+\mathbb{I}\otimes (A^T)^2]$ is the Liouvillian on the vectorized operator space, corresponding to the Lindbladian $\mathcal{L}(\cdot)=A(\cdot)A^\dagger-\frac{1}{2}\{A^\dagger A,(\cdot)\}$ with anti-commutator $\{A^\dagger A,(\cdot)\}=A^\dagger A(\cdot)+(\cdot)A^\dagger A$. 
The condition for the validity of such approximation can be obtained by expanding $\phii_A$ as $\phii_A=\mathbb{I}\otimes \mathbb{I}-\frac{\tau_A^2}{2}\hat\mcc_A^2+\frac{\tau_A^4}{24}\hat\mcc_A^4+\cdots$, and requiring that $\frac{\tau_A^2}{12}||\hat\mcc_A^2||\ll1$. Since $||\hat\mcc_A||<2||A||$, this condition can be satisfied if $\frac{\tau_A^2}{3}||A||^2\ll 1$.
%\begin{equation}
%    \phii_A=\mathbb{I}\otimes \mathbb{I}-\frac{\tau_A^2}{2}\hat\mcc_A^2+\frac{\tau_A^4}{24}\hat\mcc_A^4+\cdots.
%\end{equation}
%Then we need 
%\begin{equation}
%    \frac{\tau_A^2}{12}\abs{\abs{\frac{\hat\mcc_A^4}{\hat\mcc_A^2}}}\ll1,
%\end{equation}

Then we consider the concatenated channel
\begin{equation}
    \phii=\phii_A\phii_B\approx \phii_B+\tau_A^2\hat{ \mathcal{L}}\phii_B.
\end{equation}
Thus the channel $\phii_A$ induced by the RIM can be regarded as a perturbation acting on $\phii_B$ for small $\tau_A$. For the set of nondegenerate eigenvalues $\{v_{ij}\}$ ($i\neq j$), the eigenvalues of $\phii$ from first-order nondegenerate perturbation theory is
\begin{equation}
\begin{aligned}
        \lambda_{ij}&\approx v_{ij}+\tau_A^2\langle\langle ij| \hat{ \mathcal{L}}\phii_B |ij\rangle\rangle\\
        &\approx v_{ij}\left(1-\frac{\tau_A^2}{2}\braa{ij}\hat{\mathcal C}_A^2\kett{ij}\right).
\end{aligned}
\label{Eq:perturbEigvalOffDiag}
\end{equation}
Since that $A$ is Hermitian, the superoperator $\hat{\mathcal C}_A^\dagger = A\otimes\mathbb{I}-\mathbb{I}\otimes A^T=\hat{\mathcal C}_A$ is also Hermitian. Then the diagonal elements $\braa{ij}\hat{\mathcal C}_A^2\kett{ij}$ should be non-negative real numbers. This means that under first-order perturbation, the action of $\phii_A$ on $\phii_B$ only reduces the absolute values of $\phii$, turning them from rotation points into decaying points, without affecting their phases. Higher-order perturbation terms may bring additional small phase shifts,
%Actually, due to $\phii_A=\cos(\tau_A\hat\mcc_A)$, $\langle\langle ij|\phii_A|ij\rangle\rangle\in\mathbb{R}$, the phase drift results from the second-order perturbation, which gives that
\begin{equation}
    \lambda_{ij}=v_{ij}\left(1-\frac{\tau_A^2}{2}\braa{ij}\hat{\mathcal C}_A^2\kett{ij}+\frac{\tau_A^4}{24}\braa{ij}\hat{\mathcal C}_A^4\kett{ij}\right)+\frac{\tau_A^4}{4}\sum_{kl\neq ij}\frac{|\braa{kl}\hat\mcc_A^2\kett{ij}|^2}{v_{ij}-v_{kl}}+\cdots,
\end{equation}
where the last term with $v_{ij}-v_{kl}$ may be complex. 

However, the above analysis does not apply to the set of degenerate eigenvalues $\{v_{ii}\}$ with $v_{ii}=1$ for any $i$. According to the degenerate perturbation theory, we need to diagonalize this degenerate subspace
\begin{equation}
  \hat{ \mathcal{L}}^{(D)}=\mqty(\sum_{j\neq 1} |a_{j1}|^2 & -|a_{12}|^2&\cdots \\
  -|a_{12}|^2 & \sum_{j\neq 2}|a_{j2}|^2 & \cdots\\
  \vdots & \vdots & \ddots),
\end{equation}
in which we used $\braa{ii} \hat{ \mathcal{L}} \kett{jj}=(A^2)_{ii}\delta_{ij}-|a_{ij}|^2$ and $(A^2)_{ii}=\sum_{k} |a_{ik}|^2$ with $a_{ij}=\bra{i}A|j\rangle$. We denote the eigenvalues of $\hat{ \mathcal{L}}^{(D)}$ as $l^{(D)}_{i}\,(i=1\,...\,d)$ with $d$ being the dimension of the system, then the perturbation of the fixed points can be expressed as $\lambda_{ii}\approx1-\tau_A^2 l^{(D)}_i$. We can find that the subspace $\hat{ \mathcal{L}}^{(D)}$ is a Laplacian matrix that every row sum and column sum of it is zero. Then, there exists an eigenvalue $l^{(D)}_1=0$ with the eigenvector $\sum_{i=1}^d|ii\rangle\rangle$. This is the unit matrix in the Hilbert space and can be normalized to $\rho=\I_d/d$, which is the maximally mixed state and the fixed point of the channel. 
When $d=2$, $\hat{ \mathcal{L}}^{(D)}$ can be diagonalized easily, then we obtain $l^{(D)}_1=0$ and $l^{(D)}_2=2|a_{12}|^2$. Thus $\lambda_{11}=1$ and $\lambda_{22}\approx 1-2\tau_A^2 |a_{12}|^2$.

\subsection{Analytical and numerical verification for perturbation theory}\label{S1c}
We can analytically verify the validity of perturbation theory by the two qubit model, in which the coupling Hamiltonian is $A=g\sigma_x /2$, and $B=\omega\sigma_z/2$. Then we can write $\hat{\mathcal C}_A^2$ in the basis of $\sigma_z$
\begin{equation}
\begin{aligned}
    \hat{\mathcal C}_A^2&=(A\otimes \mathbb{I}-\mathbb{I}\otimes A^T)^2%\\
    =\frac{g^2}{2}\mqty(1&0&0&-1\\0 &1 &-1 &0\\0 &-1 &1 &0\\-1&0&0&1).
\end{aligned}
\end{equation}
According to Eq.~\eqref{Eq:perturbEigvalOffDiag}, 
\begin{equation}
\begin{aligned}
        \lambda_{12}&\approx v_{12}\left(1-\frac{\tau_A^2}{2}\braa{12}\hat{\mathcal C}_A^2\kett{12}\right)%\\
        &=\ee^{-\ii \nu}\left(1-\frac{\mu^2}{4}\right).
\end{aligned}
\end{equation}
 Similarly, $\lambda_{21}\approx \ee^{\ii  \nu}\left(1-\frac{\mu^2}{4}\right)$. Besides, $a_{12}=g/2$, then $|a_{12}|^2=g^2/4$, and we have $\lambda_{22}\approx 1-\mu^2/2$.

We then compare them with the analytically derived channel spectra in the main text, where $\lambda_{11}=1$, $\lambda_{22}=\cos(\mu)$ and $\lambda_{12,21}=\cos^2\left(\frac{\mu}{2}\right)[\cos\nu\pm\sqrt{\tan^4\left(\frac{\mu}{2}\right)-\sin^2\nu}]$.
{When $\mu\ll \nu$, we can use the Taylor expansion and take the terms up to the second order, then $\cos^2(\frac{\mu}{2})\approx 1-\frac{\mu^2}{4}$, and $\tan^4(\frac{\mu}{2})\approx\mathcal{O}(\mu^4)$. Thus $\lambda_{11}=1$, $\lambda_{12,21}=\left(1-\frac{\mu^2}{4}\right)(\cos\nu\pm \ii|\sin\nu|)\approx \ee^{\pm \ii|\nu|}\left(1-\frac{\mu^2}{4}\right)$ (here we take $-\pi<\nu\leq\pi$), and $\lambda_{22}\approx 1-\frac{\mu^2}{2}$. The eigenvalues $\lambda_{11}$ and $\lambda_{22}$ coincides with our perturbation results, while $\lambda_{12,21}$ give the same set of eigenvalues compared to the perturbation results.}

\begin{figure}
    \centering
    \includegraphics[width=0.9\linewidth]{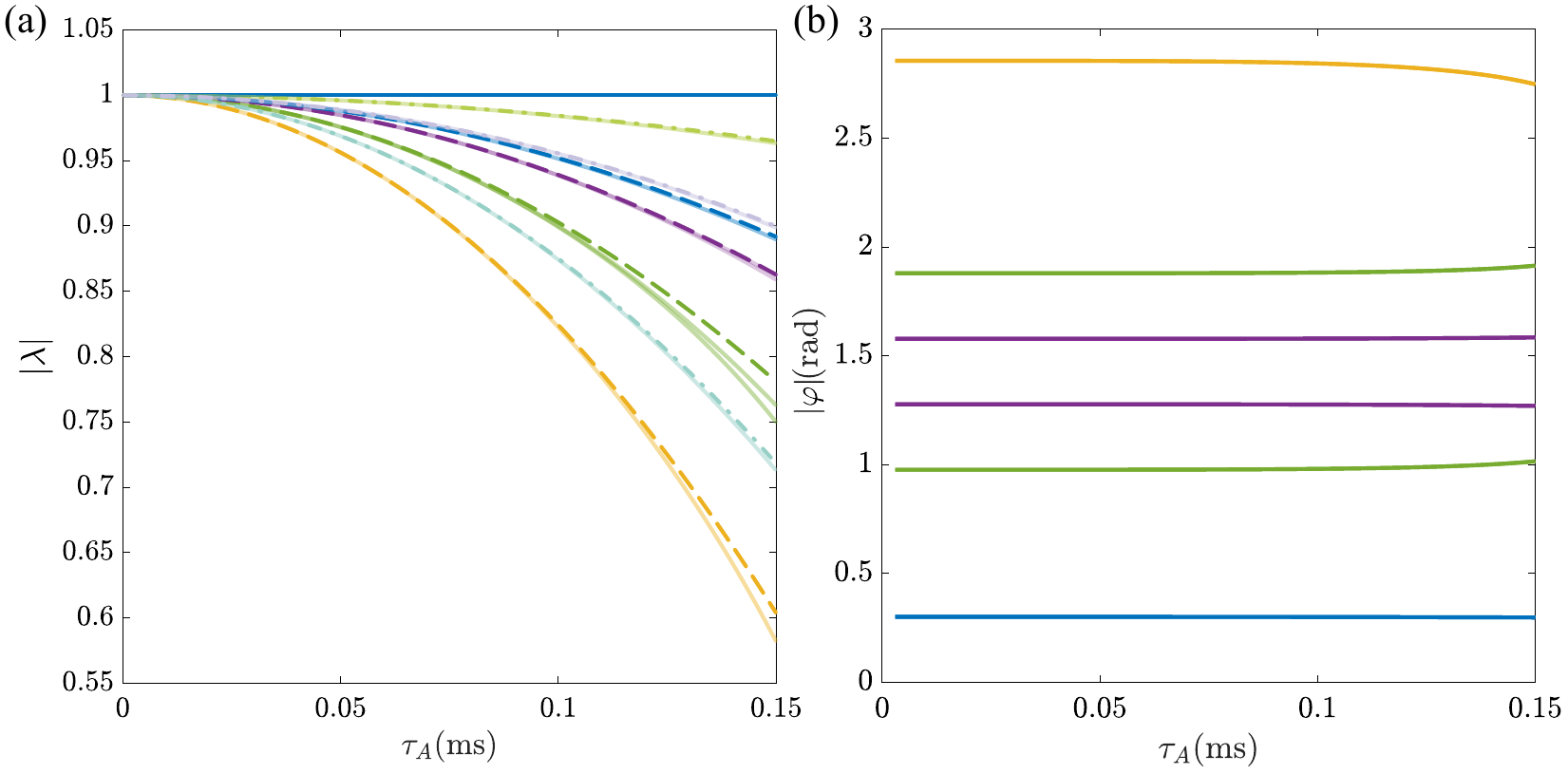}
    \caption{Numerical verification of channel perturbation theory for a two-qubit system. (a) The absolute values of the eigenvalues of $\phii$ (solid lines) as a function of $\tau_A$. The predicted eigenvalues obtained by non-degenerate perturbation (modifying on the rotating points of $\phii_B$) and degenerate perturbation (modifying on the fixed points of $\phii_B$) are shown with dashed and dot-dashed lines, respectively. (b) The phases of the eigenvalues of $\phii$ as a function of $\tau_A$.  }
    \label{Fig:numVer}
\end{figure}

%\subsection{Numerical verification}\label{S1c}
We also numerically verify the validity of the perturbation theory by the two-qubit system in Fig.~\ref{Fig:numVer}. Here we take the same model used in Example II of the main text, i.e.,  $A=\sum_{n=1}^2 \vb*{h}_n\cdot \vb*{I}_n$ and $B\approx\omega\sum_{n=1}^2 I^z_i+D(I_1^+I_2^-+I_1^-I_2^+-4I_1^z I_2^z)$, where $\vb*{h}_n=(h_n^x, h_n^y, h_n^z)$ is hyperfine coupling parameter, $\vb*{I}_n=(I_n^x, I_n^y, I_n^z)$ is the $n$th nuclear spin operator ($I_n^i=\sigma_n^i/2$) and $I_n^{\pm}=I^x_n\pm i I^y_n$.  We can see that the predicted eigenvalues coincide well with the true eigenvalues, and the phases are almost stable under the perturbation.

\subsection{Amplitudes of the weak-measurement signals}\label{sec:amp}
Then we discuss the amplitudes of the weak-measurement signals, determined by the coefficient $c_{ij}$ corresponding to the eigenvalue $\lambda_{ij}$ that contributes to the signal $f_1$. According to the main text, $c_{ij}=\Tr(M_1^\dagger M_1 R_{ij})\Tr(L_{ij}^\dagger\rho)$. When $\tau_A||A||$ is small, the eigenstates of $\Phi$ can be taken as those of $\Phi_B$, which are $\{R_{ij}=L_{ij}=|i\rangle\langle j|\}_{i,j=1}^d$. We have
\begin{equation}
\begin{aligned}
        M_1^\dagger M_1&=\frac{1}{4}(U_0^\dagger+\ee^{-\ii \phi} U_1^\dagger)(U_0+\ee^{\ii  \phi} U_1)%\\
        =\frac{1}{4}(2\I+\ee^{-\ii \phi}U_1^\dagger U_0+\ee^{\ii  \phi} U_0^\dagger U_1).
\end{aligned}
\end{equation}
Since $\Tr(R_{ij})=0$ for $i\neq j$, we have
%for the rotating points (which we want to estimate)
\begin{equation}
\begin{aligned}
        c_{ij}&\propto\Tr(\frac{\ee^{-\ii \phi}U_1^\dagger U_0+\ee^{\ii  \phi} U_0^\dagger U_1}{2}\ket i\bra j)\\
        &=\Tr(\frac{\ee^{-\ii \phi}\ee^{-2\ii A\tau_A}+\ee^{\ii  \phi} \ee^{2\ii A\tau_A}}{2}\ket i\bra j)\\
        %&=\Tr[ \cos(2A\tau_A+\phi)\ket i\bra j]\\
        %&=\bra{j}\cos(2A\tau_A+\phi)\ket i\\
        &=\sum_{k}\cos(2a_{k}\tau_A+\phi)\braket{j}{a_k}\braket{a_k}{i},
\end{aligned}
\end{equation}
where the eigenbasis of $A$ is denoted as $\{|a_k\rangle\}_{k=1}^d$, i.e., $A=\sum_{k}a_k|a_k\rangle\langle a_k|$. Then when $\tau_A$ is small, we have 
\begin{equation}
    \cos(2a_{k}\tau_A+\phi)\approx\cos\phi- (2a_k\tau_A)\sin\phi- {2(a_k\tau_A)^2}\cos\phi,
\end{equation}
so
\begin{equation}
 \begin{aligned}
    c_{ij}&\propto\sum_{k}[\cos\phi- 2a_k\tau_A\sin\phi- {2(a_k\tau_A)^2}\cos\phi]\braket{j}{a_k}\braket{a_k}{i}\\
    &= -2\tau_A\sin\phi\sum_{k}a_k\braket{j}{a_k}\braket{a_k}{i}- {2\tau_A^2}\cos\phi\sum_{k}a_k^2\braket{j} {a_k}\braket{a_k}{i}\\
    &=-2[\tau_A\sin\phi \langle j| A|i\rangle+\tau_A^2\cos\phi \langle j|A^2|i\rangle].
\end{aligned}
\label{Eq:cij}
\end{equation}
where we have used $\sum_{k}\braket{j}{a_k}\braket{a_k}{i}=\delta_{ij}$. This clearly shows the signal amplitude grows with $\tau_A.$

Our method tracks the evolution of the target system under the action of sequential quantum channels, which is similar to the analysis of the correlation function mentioned in Sec.~\ref{Sec:corrFunc} \cite{pfender2019} or the evolution of measurement results for a single qubit target system with $A=g\sigma_x/2$ and $B=\omega\sigma_z/2$  \cite{cujia2019}. 
%which is the same as our method. In that work, the theoretical tool is analytically

In Ref.~\cite{cujia2019}, the measurement results in each round of observation is analyzed by directly solving the evolution of the spin-1/2 nuclear quantum state. By taking $\phi=\frac{\pi}{2}$, they find that the signal amplitude grows linearly with $\tau_A$, i.e., $c_j\propto \tau_A$ while the decoherence grows quadratically, i.e., $|\lambda_{ij}|\propto \tau_A^2$. We note that these results are consistent with ours [see Eqs.~\eqref{Eq:perturbEigvalOffDiag} and \eqref{Eq:cij}]. 
While the methods in these works are hard to analyze multi-spin systems, our framework based on quantum channels can be easily extended to any complex target quantum system.

%as we can see, those works require a fully solvable model, which is hard to extend to multi-spin systems. The theoretical model we propose is appropriate for Hamiltonian learning applicable to any system.

%then the lowest two orders of $c_{ij}$ are 
%\begin{equation}
%    \begin{aligned}
%        c_{ij}&\propto -2\tau_A\sin\phi\sum_{k}a_k\braket{j}{a_k}\braket{a_k}{i}- {2\tau_A^2}\cos\phi\sum_{k}a_k^2\braket{j}{a_k}\braket{a_k}{i}\\
%        &=-2[\tau_A\sin\phi \langle j| A|i\rangle+\tau_A^2\cos\phi \langle j|(A^2)|i\rangle],
%    \end{aligned}
%\end{equation}

% \subsection{Relation with correlation measurements}

\subsection{Effect of the free Hamiltonian in RIM}\label{sec:freeHam}
When we include the free Hamiltonian $B$ in the evolution Hamiltonian in RIM, i.e., $H=\sigma_{\q}^z\otimes A+\mathbb{I}_{\q}\otimes B$, the propagators become $U_0=\ee^{-\ii (A+B)\tau_A}$ and $U_1=\ee^{-\ii (-A+B)\tau_A}$. The propagator $U_0$ can be considered as being generated by a Hamiltonian $H_0=A+B$, then we transform into an interaction picture with a free Hamiltonian $B$ and the interaction part $A$. We can define $A_I(t)=\ee^{\ii  Bt}A\ee^{-\ii Bt}$, then in the interaction picture, $U_0=\ee^{-\ii B\tau_A}U_I(t)$, where $\ii\dv{U_I(t)}{t}=U_I(t)A_{I}(t)$. We can write $U_I(t)=\mct \ee^{-\ii \int_{0}^{\tau_A} A_I(t)\dd t}$ with $\mct$ being the time-ordering operator, and thus
\begin{equation}
    U_0=\ee^{-\ii B\tau_A}\mct \ee^{-\ii \int_{0}^{\tau_A} A_I(t)\dd t}=\tilde\mct \ee^{-\ii \int_{0}^{\tau_A} \tilde A_I(t)\dd t}\ee^{-\ii B\tau_A}=\tilde U_0 U_B,
\end{equation}
where $\tilde \mct$ is an anti-time-ordering operator, $\tilde A_I(t)=\ee^{-\ii Bt}A\ee^{\ii  Bt}$, $U_B=\ee^{-\ii B\tau_A}$ and 
$\tilde U_0= \tilde{\mathcal T}\ee^{-\ii \int_0^{\tau_A} \tilde A_I(t)\dd t}$. We use the second form in the derivation since $\phii_A$ acts after $\phii_B$, and through this we can combine the unitary channel $U_B\otimes U_B^*$ with $\phii_B$. When $\tau_A$ is small, we have
\begin{equation}
   \tilde U_0
= \mathbb{I}-\ii\int_0^{\tau_A} \dd t_1\tilde A_I(t_1)
-\int_0^{\tau_A} \dd t_1 \int_0^{t_1} \dd t_2 \tilde A_I(t_2)\tilde A_I(t_1)+\mathcal O({\tau_A}^3),
\end{equation}
%where $ \tilde U_0= \tilde{\mathcal T}\ee^{-\ii \int_0^{\tau_A} \tilde A_I(t)\dd t}$.
when $\tau_A ||B||$ is small, we can expand $\tilde A_I(t)$ as
\begin{equation}
    \tilde A_I(t)=\ee^{-\ii Bt}A\ee^{\ii  Bt}
=A+\ii t[A,B]+\mco(t^2),
\end{equation}
we have 
\begin{equation}
    \ii\int_0^{\tau_A} \dd t_1\tilde A_I(t_1)=\ii\tau_A A-\frac{{\tau_A}^2}{2}[A,B],
\end{equation}
and
\begin{equation}
    \int_0^{\tau_A}\dd t_1\int_0^{t_1}\dd t_2 \tilde A_I(t_2)\tilde A_I(t_1)
= \frac{{\tau_A}^2}{2}A^2+\mco({\tau_A}^3),
\end{equation}
then up to the second order of $\tau_A$, we have
\begin{equation}
    \tilde U_0=\mathbb{I}-\ii\tau_A A-\frac{{\tau_A}^2}{2}(A^2-[A,B]).
\end{equation}
Similarly, we can also obtain $U_1=\tilde U_1 U_B$ with
\begin{equation}
    \tilde U_1=\mathbb{I}+\ii\tau_A A-\frac{{\tau_A}^2}{2}(A^2+[A,B]).
\end{equation}
 %The equations above can also be derived from the Zassenhaus formula.
Thus we have 
\begin{equation}
    \tilde U_0\otimes \tilde U_0^*\approx\mathbb{I}\otimes \mathbb{I}-\ii\tau_A(A\otimes \mathbb{I}+\mathbb{I}\otimes A^T)+ \tau_A^2 \left[A\otimes A^T 
-\frac{1}{2}(A^2-[A,B])\otimes \mathbb{I} - \frac{1}{2} \mathbb{I}\otimes((A^T)^2 - [A^T,B^T])\right],
\end{equation}
and
\begin{equation}
   \tilde U_1\otimes \tilde U_1^*\approx\mathbb{I}\otimes \mathbb{I}+\ii\tau_A(A\otimes \mathbb{I}+\mathbb{I}\otimes A^T)+ \tau_A^2 \left[A\otimes A^T 
-\frac{1}{2}(A^2+[A,B])\otimes \mathbb{I} - \frac{1}{2} \mathbb{I}\otimes((A^T)^2 + [A^T,B^T])\right],
\end{equation}
resulting in
\begin{equation}
\begin{aligned}
        \phii_A&\approx\frac{1}{2}(\tilde U_0\otimes \tilde U_0^*+\tilde U_1\otimes \tilde U_1^*)(U_B\otimes U_B^*)\\
        &=(\mathbb{I}+\tau_A^2\hat{L})(U_B\otimes U_B^*),
\end{aligned}
\end{equation}
%where $U_B=\ee^{-\ii B\tau_A}$.

\begin{figure}
    \centering
    \includegraphics[width=0.5\linewidth]{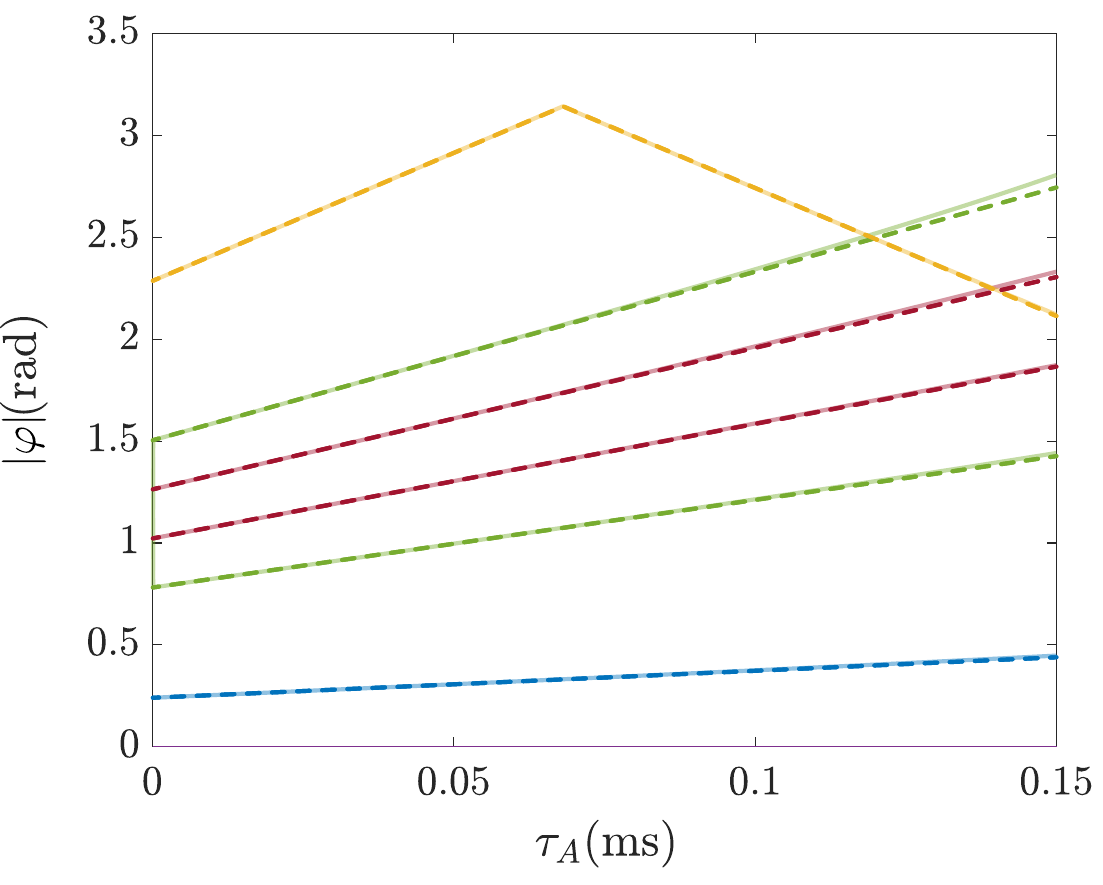}
    \caption{The phases of eigenvalues of quantum channel $\phii$ (lines) and $\tilde \phii_B$ (dashed lines) as functions of RIM evolution time $\tau_A$. }
    \label{fig:freeHamiltonian}
    \end{figure}

One can see that the effect of free Hamiltonian during the RIM channel can be understood as simply appending a period of free evolution to the original channel, and this free evolution can further be absorbed into $\phii_B$,
\begin{equation}
    \phii=\phii_A\phii_B=(\mathbb{I}+\tau_A^2\hat{ \mathcal{L}})\tilde\phii_B,
\end{equation}
with $\tilde\phii_B=\ee^{-\ii B(\tau_A+\tau_B)}\otimes \ee^{\ii  B^T(\tau_A+\tau_B)}$.

Below we show a numerical simulation to illustrate the property of this channel. We use the two-qubit model introduced in the main text and Sec.~\ref{S1c}, and change the evolution Hamiltonian from $H=\sigma_{\q}^z\otimes A$ to $H=\sigma_{\q}^z\otimes A+\mathbb{I}_{\q}\otimes B$. As we can see in Fig. \ref{fig:freeHamiltonian}, the phases of eigenvalues of $\phii$ match well with those of $\tilde \phii_B$, showing that we can still estimate the parameters of the Hamiltonian.
%The evolution Hamiltonian in the RIM is 
% \begin{equation}
%     H=\sigma_{\q}^z\otimes \sum_{n=1}^2 \vb*{h}_n\cdot \vb*{I}_n+\mathbb{I}_{\q}\otimes\left[ \omega\sum_{n=1}^2 I^z_i+D\sum_{n=1}^2(I_1^+I_2^-+I_1^-I_2^+-4I_1^z I_2^z)\right],
% \end{equation}

\subsection{Examples of detecting nuclear spin clusters containing multiple spins}\label{Sec:threeQ}

\begin{figure}
    \centering
    \includegraphics[width=\linewidth]{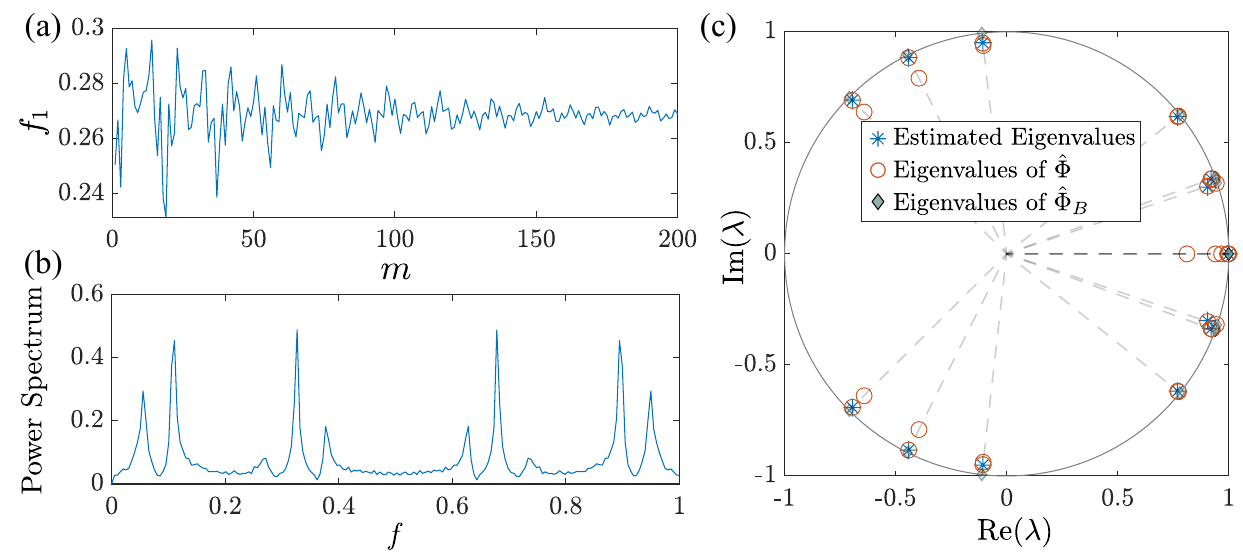}
    \caption{Hamiltonian parameter estimation for a three-spin target system. (a) Signal and (b) Fourier transformed spectrum. (c) The estimated channel eigenvalues (blue stars) and the eigenvalues of $\phii$ (red circles) and $\phii_B$ (blue diamonds). Parameters are $h_1/2\pi=26.6$ kHz, $h_2/2\pi=32.2$ kHz, $h_3/2\pi=49.4$ kHz, $D_{12}/2\pi=475.6$ Hz, $D_{13}/2\pi=238.3$ Hz, $D_{23}/2\pi=352.4$ Hz, $\omega_L/2\pi=110.3$ kHz, $\tau_A=0.955\,\,\mu$s and $\tau_B=906.9\,\,\mu$s.}
    \label{fig:threespin}
\end{figure}

\begin{table}[htbp]
\caption{Estimated phases and the corresponding Hamiltonian parameters for a three-spin cluster. The estimated values of parameters are $D_{12}/2\pi=475.4$ Hz, $D_{13}/2\pi=239.2$ Hz, $D_{23}/2\pi=351.7$ Hz, with the estimation errors being 0.06\%, 0.38\% and 0.18\%, respectively.
%the relevant errors are $\delta(\omega)=0.3\%$ and $\delta(D)=0.8\%$
}
\begin{ruledtabular}
\begin{tabular}{lll}
Phases ($^\circ$) & $\tilde \bij/2\pi$ (Hz)  &  Parameters \\
\hline
18.03 & 110.5 &$D_{23}-D_{13}$ \\
19.94 & 122.1 & $D_{12}-D_{23}$\\
38.76 & 237.4 & $D_{12}-D_{13}$ \\
96.43 & 590.7 & $D_{13}+D_{23}$ \\
116.6 & 714.0 &$D_{12}+D_{13}$\\
135.1 & 827.9 & $D_{12}+D_{23}$ \\
\end{tabular}
\end{ruledtabular}\label{Table}
\end{table}

Here we show an example of Hamiltonian parameter estimation for a three-spin target system in Fig.~\ref{fig:threespin}. With a strong magnetic field, the coupling and free Hamiltonian are
\begin{equation}\label{Eq:mulNuc}
    A=\sum_{k=1}^K \vb*h^{(k)}\cdot\vb*I^{(k)},\quad B=\omega_L\sum_{k=1}^K I_z^{(k)}+\sum_{j<k}D_{jk}I_z^{(j)}I_z^{(k)},
\end{equation}
where $K$ is the number of nuclear spin, $\vb*h^{(k)}=(h_{x}^{(k)},h_{y}^{(k)},h_{z}^{(k)})$ is the hyperfine coupling parameter between the ancilla and the $k$th nuclear spin, $\vb*I^{(k)}=(I^{(k)}_x,I^{(k)}_y,I^{(k)}_z)$ is the nuclear spin operator for the $k$th nuclear spin, $\omega_L=\gamma_n B$ is the Larmor frequency with $\gamma_n$ and $B$ being the gyromagnetic ratio and magnetic field strength respectively, and $D_{jk}$ is the coupling strength between the $j$th and the $k$th spin. We consider three nuclear spins here, i.e., $K=3$. We suppose that the Larmor frequency is known and our objective is to obtain the coupling parameters.  

Then the eigenvalues of $B$ are
\begin{equation}
    b_{\alpha,\beta,\gamma}=\frac{\omega_L}{2}(\alpha+\beta+\gamma)+\frac{1}{4}\left(D_{12}\alpha\beta+D_{13}\alpha\gamma+D_{23}\beta\gamma\right),
\end{equation}
with $\alpha,\beta,\gamma\in\{1,-1\}$. Then the channel has 64 eigenvalues related to $\bij=\pm\{\pm (D_{12}- D_{23}),\pm (D_{12}- D_{13}),\pm (D_{13}- D_{23}),2\omega_L\pm (D_{12}\pm D_{23}),2\omega_L\pm (D_{12}\pm D_{13}),2\omega_L\pm (D_{13}\pm D_{23}),4\omega_L\pm (D_{12}+ D_{23}),4\omega_L\pm (D_{12}+ D_{13}),4\omega_L\pm (D_{13}+ D_{23}),\pm 6\omega_L,0\}/2$. We cannot extract eigenvalues from such a dense spectrum.
Since that $\omega_L\gg D$ ($|\omega_L|\sim 10^3 |D|$ here), we choose $\tau_B=2\pi q/\omega_L$ with $q\in\mathbb Z$ to extract the coupling parameters $D$. Through this method, we are also able to reduce the number of eigenvalues to just thirteen distinct effective $\tilde\beta_{ij}$ (in $\tilde \beta_{ij}$ we omit the term of $\omega_L$ since $\omega_L\tau_B=0\!\!\mod2\pi$), where $\tilde \beta_{ij}=\{0,\pm (D_{12}\pm D_{23}),\pm (D_{12}\pm D_{13}),\pm (D_{13}\pm D_{23})\}/2$. Then the coupling parameters can be extracted by the enumeration method, as listed in Table.~\ref{Table}.%, however, the parameters may be allocated in an error sequence without a priori knowledge.

We note that the channel spectrum for a larger spin cluster can be very dense, making it difficult to extract the whole spectrum and all the Hamiltonian parameters by the MP method. We expect that the scheme with additional dynamical decoupling sequences during RIM evolution period may eliminate the unwanted noise and focus on some specific coupling parameters \cite{Taminiau2012,Abobeih2019,Bradley2019}. Moreover, it is also interesting to develop more advanced spectral analysis methods (e.g., with some machine learning algorithms).

\section{Performance analysis}

\subsection{Sample complexity}
In this paper, we use the frequency $f_1$ to estimate the probability $p_1$. Here we analyze the accuracy of tracking under limited sample number. According to the Hoeffding's inequality
\begin{equation}
    \Pr(|p_1^{(m)}-f_1^{(m)}|\geq\delta)\leq 2\ee^{-2N_s\delta^2}.
\end{equation}
Then if we want to obtain a precision $\delta$ with probability $1-\epsilon$, the minimum sample number is
\begin{equation}
    N_s=\frac{1}{2\delta^2}\ln(\frac{2}{\epsilon}).
\end{equation}
Considering that the amplitudes of the signals in the examples we take are in the magnitude order of $10^{-2}\sim 10^{-1}$, then we choose $\delta= 10^{-3}$. Under this choice, the required sample size $N_s$ is on the order of $10^6$.

\subsection{Effect of RIM duration}
Then we briefly discuss the impact of coupling duration $\tau_A$ on the accuracy. As illustrated in Sec.~\ref{sec:amp}, the signal amplitude grows with $\tau_A$ from zero, then when $\tau_A$ is too small, it is difficult to attain a sufficient signal-to-noise ratio to extract the channel spectrum. While if $\tau_A$ is too large, the higher order perturbation may tune the frequencies of the eigenvalues, resulting in the decreasing of the accuracy.
In the system in which the free Hamiltonian is inevitably included in the RIM, we should also consider the impact of the higher-order Dyson series as the growth of $\tau_A$, which may induce additional errors.

\begin{figure}
    \centering
    \includegraphics[width=\linewidth]{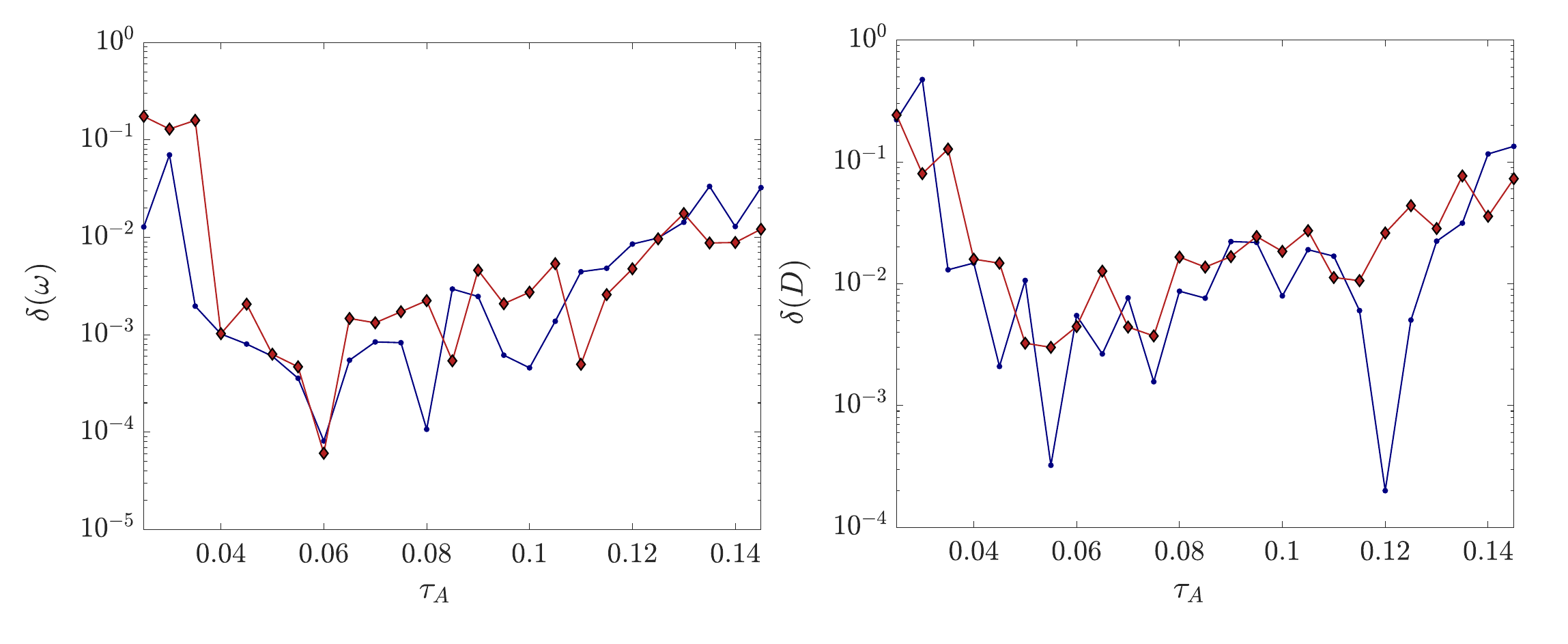}
    \caption{Estimation errors of Larmor frequency $\omega$ and coupling strength $D$ as functions of coupling duration $\tau_A$. The accuracy in the case that free Hamiltonian is included (not included) in the RIM is shown by the blue (red) line. The model and parameters are the same as those in Fig.~2 in the main text. }
    \label{fig:coupling}
\end{figure}

We perform the numerical simulation in Fig.~\ref{fig:coupling} for the two-qubit system in the main text  (i.e.,  $A=\sum_{n=1}^2 \vb*{h}_n\cdot \vb*{I}_n$ and $B\approx\omega\sum_{n=1}^2 I^z_i+D(I_1^+I_2^-+I_1^-I_2^+-4I_1^z I_2^z)$). Here we compare the accuracy of free evolution being included (blue line) and excluded (red line) from RIM. In the weak-measurement limit, we can see that the estimation accuracies are similar in the two cases, and the accuracy first increases due to the growth of signal amplitude and then decreases due to the altered frequencies.

\subsection{Effect of target system noise}\label{sec:noise}
\begin{figure}
    \centering
    \includegraphics[width=\linewidth]{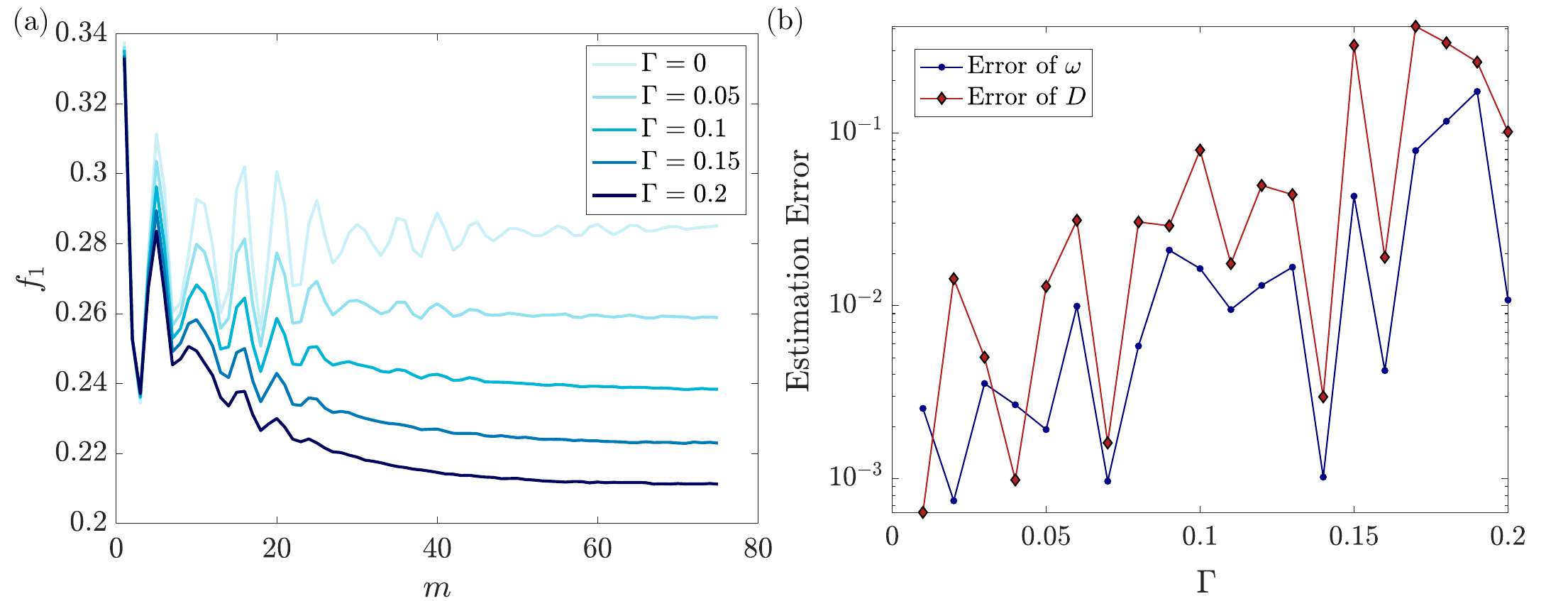}    
    \caption{Hamiltonian parameter estimation with the Lindblad noise.  (a) The signal with different dissipation strength. (b) The estimation error of $\omega$ and $D$ as functions of dissipation strength $\Gamma$. The model and parameters are the same as those in Fig.~2 in the main text.
 }
    \label{Fig: Lindblad}
\end{figure}

Here we analyze the error of Hamiltonian parameter estimation with the presence of Lindblad noise. The evolution of the system can be described by the Lindblad master equation
\begin{equation}
    \dv{\rho_{\rm tot}}{t}=-i[H,\rho_{\rm tot}]+\sum_k \Gamma_k\left(L_k\rho_{\rm tot} L_k^\dagger-\frac{1}{2}\{L_k^\dagger L_k,\rho_{\rm tot}\}\right),
\end{equation}
where $\rho_{\rm tot}$ is the state of the composite system. Here we simulate the two-qubit target system in the main text where $L_1=\sigma_{\q}^z$ and $L_2=\sigma_{\q}^-=\ket{1}_{\q}\bra{0}$ denote probe dephasing and relaxation, $L_3=\sigma_z^{(1)}$, $L_4=\sigma_-^{(1)}$, $L_5=\sigma_z^{(2)}$ and $L_6=\sigma_-^{(2)}$ denote the dephasing and relaxation of each target spin, and we set the same dissipation strength $\Gamma$ for each $L_k$. We note that this type of noise does not tune the frequency of the eigenvalues. However, the noise induces additional damping on them, resulting in a decrease in the signal amplitude [Fig.~\ref{Fig: Lindblad} \blue{(a)}], which makes it difficult to extract the frequencies with limited sample numbers. We show the simulation in Fig.~\ref{Fig: Lindblad} \blue{(b)}, where we can see that with the growth of the dissipation strength $\Gamma$, the estimation error also tends to increase because the amplitudes of some frequencies are too small to estimate within a finite sample number. Due to the limited sample numbers, the figure also exhibits some fluctuations.

\subsection{Effect of target system size}

\begin{figure}
    \centering
    \includegraphics[width=\linewidth]{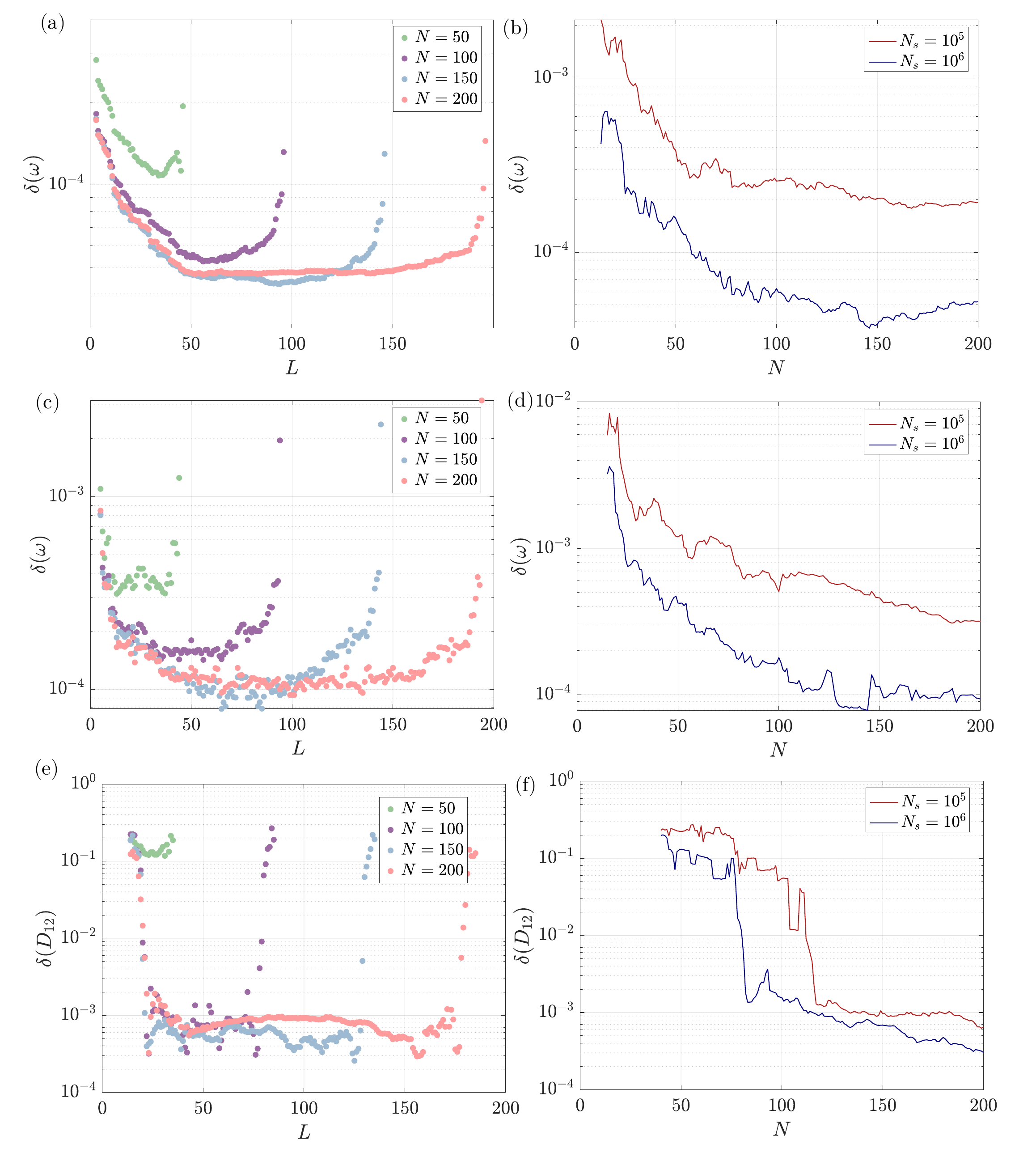}
    \caption{Estimation error as functions of (left) pencil length $L$ and (right) sequence length $N$ for the (a,b) one, (c,d) two and (e,f) three nuclear spin bath. We choose $N_s=10^6$ for (a,c,e) and $L=N/2$ for (b,d,f).}
    \label{fig:perfOne}
\end{figure}

Here we use the model shown in Eq.~\eqref{Eq:mulNuc} to illustrate the accuracy of Hamiltonian parameter estimation in the system with $K=1,2,3$. 
% \subsubsection{Case 1: $K=1$}
% For the target system containing just one nuclear spin, we can estimate the Larmor frequency $\omega$ of the spin [Fig.~\ref{fig:perfOne}]. While for $K=2$, we aim to estimate both $\omega$ and the coupling strength $D$ [Fig.~\ref{fig:perfTwo}]. The task here for $K=3$ is the same with that in Sec.~\ref{Sec:threeQ}, where we estimate the three coupling strength [Fig.~\ref{fig:perfThree}]. 
Here we investigate the estimation error of Larmor frequency for the case of $K=1,2$ while the error of $D_{12}$ for $K=3$.  We choose different pencil length $L$, sequence length $N$ and sample number $N_s$ to compare the performance. We note that the free evolution Hamiltonian is also included in the RIM and the result is summarized in Fig.~\ref{fig:perfOne}.  In general, precision tends to increase with $N$ and $N_s$. The  precision is approximately higher in the range $L\in (L/4, 3L/4)$. 
 As the number of spins increases, estimation accuracy decreases. %With a large number of spins and a small sample size, this can result in the failure to readout all phases or incorrect phase readings, leading to significant errors.

%\subsubsection{Case 2: $K=2$}
% \begin{figure}
%     \centering
%     \includegraphics[width=\linewidth]{FigS7.pdf}
%     \caption{Estimation error of (a-b) Larmor frequency $\omega$ and (c-d) coupling strength $D$ as functions of pencil length $L$ and sequence length $N$ for a spin cluster containing two nuclear spins. We choose $N_s=10^6$ for (a,c) and $L=N/2$ for (b,d).}
%     \label{fig:perfTwo}
% \end{figure}
% \begin{figure}
%     \centering
%     \includegraphics[width=\linewidth]{FigS8.pdf}
%     \caption{Estimation error of different coupling strength $D_{k_1k_2}$ ($k_1,k_2\in\{1,2,3\}$) as functions of pencil length $L$ and sequence length $N$ for a spin cluster containing three nuclear spins. We choose $N_s=10^6$ for (a,c,e) and $L=N/2$ for (b,d,f).}
%     \label{fig:perfThree}
% \end{figure}

%\newpage
\bibliography{NHP}